\documentclass[galaxies,article,accept,pdftex,moreauthors]{Definitions/mdpi} 

\firstpage{1} 
\pubvolume{1}
\issuenum{1}
\articlenumber{0}
\pubyear{2025}
\copyrightyear{2025}
\externaleditor{Firstname Lastname}
\datereceived{14 February 2025 } 
\daterevised{14 March 2025 } 
\dateaccepted{21 March 2025 } 
\datepublished{ } 
\hreflink{https://doi.org/} 

\Title{A Comprehensive Analysis on the Nature of the Spiral Arms in NGC~3686, NGC~4321, and NGC~2403}

\TitleCitation{A Comprehensive Analysis on the Nature of the Spiral Arms in NGC~3686, NGC~4321, and NGC~2403}

\Author{Valeria Kostiuk 
 $^{1,}$*\orcidA{}, Alexander Marchuk $^{2,3}$\orcidB{}, Alexander Gusev $^{1}$\orcidC{} and Ilia V. Chugunov 
 $^{2}$}

\AuthorNames{Valeria Kostiuk, Alexander Marchuk, Alexander Gusev and Ilia Chugunov}

\AuthorCitation{Kostiuk, V.; Marchuk, A.; Gusev, A.; Chugunov, I.}

\address{%
$^{1}$ \quad Sternberg Astronomical Institute, Lomonosov Moscow State University, Universitetsky pr. 13, 119234~Moscow, Russia; gusev@sai.msu.ru\\
$^{2}$ \quad Central (Pulkovo) Astronomical Observatory, Russian Academy of Sciences, Pulkovskoye chaussee 65/1, 196140 St.~Petersburg, Russia; astro@gmail.com (A.M.); chugunov21@list.ru (I.C.)\\
$^{3}$ \quad Saint Petersburg State University, The Faculty of Mathematics and Mechanics, 
 Universitetsky pr. 28, 198504 St. Petersburg, Russia}

\corres{Correspondence: valeriekostiuk@yandex.ru}

\abstract{In theoretical investigations, various mechanisms have been put forward to explain the emergence of spiral patterns in galaxies. One of the few ways to find out the nature of spirals in a particular galaxy is to consider the so-called corotation radius, or corotation resonance. A distinctly defined corotation resonance is likely to indicate the existence of a spiral density wave, while the chaotic distribution of their positions may suggest a dynamic nature to the spiral structure. In this study, we analyzed measurements of the corotation radius obtained using several methods for three galaxies (NGC 3686, NGC 4321, and NGC 2403) that exhibit different morphologies of spiral structures. We also performed independent measurements to estimate the location of the resonance, which allowed us to determine whether each galaxy has a clear corotation radius position. This examination, along with other tests such as stellar age gradient, interlocking resonances, and the radial distribution of metallicity, enables us to understand the mechanism that may be responsible for the formation of spiral arms in the studied galaxies.}

\begin{document}


\section{Introduction}
Since galaxies began to be distinguished by their spiral structures, it became interesting to find out what could be the reasons for this clear difference in their appearance. A~simple classification of spiral galaxies divides them into three types~\cite{Elmegreen1982,Elmegreen1987,Elmegreen2011,Buta2015}: ``flocculent'', containing many short arms, ``grand-design'', exhibiting two distinct spiral arms, and~``multi-armed'' with three or more moderately long arms. There have been many attempts to link the basic properties of galaxies with the features of their spiral structure, such as their strength, the~pitch angle (which measures the tightness of the spiral), and~the number of arms. In~particular, the~authors of~\cite{Hart2016} demonstrated that the proportion of galaxies with more than three arms increases with the stellar mass. It was also found that the pitch angle correlates with the size of the bulge~\cite{Freeman1970, Davis2015}, the~velocity dispersion in the central region~\cite{YuHo2019}, and~the mass of a supermassive black hole~\cite{Seigar2008}. In~addition, the~authors of~\cite{Kendall2015} noted a relation between the strength of spirals and tidal forces from nearby companion galaxies. In~addition, it has been shown that flocculent galaxies tend to have lower masses~\cite{Elmegreen2011} and lower rotational velocities~\cite{Sarkar2023} rather than grand-design galaxies. Since it has become possible to analyze the properties of galaxies using a significantly larger sample, some of the facts mentioned above have been re-examined and clarified. For~example, the~connection between the tightness of a galaxy's spiral arm and the prominence of its bulge has become uncertain (for more details, see~\cite{Masters2019}).
\par
All the findings discussed above allow us to understand and estimate the influence of galactic properties on the spiral structure. This difference in spiral morphology may also be related to the mechanisms responsible for their formation, however this question is still under debate. From~a theoretical point of view, there are two main mechanisms that can explain the nature of spiral structures. The~first suggests the existence of a quasi-stationary density wave~\cite{Lin1964,Lin1967,Roberts1969,Bertin1989}, which can explain active star formation in spiral arms as a result of gas compression due to the propagation of the density wave. The~second idea is that spiral structures are caused by transient and recurring instabilities, leading to ``dynamic'' spiral arms that appear and disappear in cycles~\cite{Julian1966,Sellwood1984,Sellwood2011}. This model of local gravitational instability suggests that star formation is stochastic. Besides, these two scenarios have difference in pattern speed profile, constant for the density wave and decreases with radius for dynamic spirals. A~grand-design spiral arm is commonly referred to the density wave scenario, since it was shown by~\cite{Lin1967,Thomasson1990} that spiral modes are stable only if the number of spirals is less than three. There were many attempts to create spiral structures in simulations, and~it has been shown that the formation of a steady two-armed pattern is only possible under special conditions~\cite{Toomre1969, Thomasson1990, Donner1994, D'Onghia2013}. In~contrast, the~dynamic spiral arm scenario is believed to produce a pattern that resembles multi-armed or flocculent galaxies. Besides, the~formation of a grand-design pattern is also attributed to the density wave driven by a bar perturbation~\cite{Sanders1976,Sellwood1988}. However, the~question of a direct relationship between bar and spiral strength, or~between their pattern speeds, remains unclear. Additionally, the~generation of a steady spiral pattern may occur due to tidal interaction with a companion~\cite{Kormendy1979,Byrd1992}. Although~spiral formation driven by a companion or bar is common in numerical simulations~\cite{Rautiainen1999,Oh2008,Kumar2022,Struck2011,Dobbs2010,Semczuk2017}, the~question whether these perturbations are sufficient to induce a spiral density wave in real galaxies remains open.    
\par
The other scenario predicts the existence of multiple modes (patterns) instead of a single one. In~this case, each pattern rotates at an individual pattern speed, forming co-called ``mode-coupling'', which was mathematically and numerically studied in~\cite{Sygnet1988,Masset1997}. Furthermore, simulations by~\cite{Rautiainen1999} showed that these modes can be coupled by resonances between the inner and outer structures, which leads to an effective transfer of angular momentum outward the disk. The~existence of galaxies with multiple patterns has been demonstrated in numerous examples of real objects in~\cite{Buta&Zhang2009, Meidt2009, Font2014a,Marchuk2024c,Kostiuk2024}. Furthermore, as~noted in~\cite{Font2014a}, it is possible that the number of such galaxies could increase if some objects are investigated based on mode-coupling features rather than focusing on evidence supporting the density wave scenario.
\par
As was sufficiently explained in the review~\cite{Dobbs2010}, the~role of each of these mentioned mechanisms in generating the observed spiral patterns in galaxies remains a subject of controversy. To~investigate this issue, it is crucial to focus on real spiral galaxies. To~distinguish between spiral arms formed by different mechanisms, there are several observational tests available. The~first method is based on measuring the pattern speed directly, which can be done using the Tremaine-Weinberg (``T-W'') method~\cite{Tremaine1984}. This method analyzes the distributions of density and velocity of both the gaseous and stellar components of a galaxy through the slits parallel to the major axis. However, this approach initially assumes a fixed angular velocity for the galaxy, which may not be true for all objects. For~instance, applying a modified version of this method~\cite{Merrifield2006}, which account radial variations in the pattern speed, has shown that some galaxies may have multiple patterns rotating at different angular velocities~\cite{Meidt2009}. As~a result, this method can be ambiguous and should be used with caution. 
\par
Another way to determine whether a galaxy has a fixed pattern speed is by considering the so-called corotation radius or resonance (CR). At~this radius, the~angular velocity of the spiral equals that of the disk. If~a galaxy has a constant pattern speed, it can be divided into two parts: inside the corotation circle, the~disk material rotates faster than the spiral, and~vice~versa outside. Thus, in~the central region of the galaxy, the~gaseous clouds contained in the disk overtake the spiral pattern and are compressed by the spiral density wave, triggering the star-formation process at the inner edge of the spiral arm. Newly formed stars then begin to evolve as they cross the spiral arm and move towards the direction of galactic rotation. Therefore, within~the CR, we should observe a stellar age gradient from upstream to downstream across the pattern, and~in the opposite direction outside. This is another strong observational evidence that supports the density wave theory. However, despite the apparent simplicity of this argument, studies using different tracers of stellar populations or methods to detect age gradients have yielded contradictory results~\cite{Vallee2020}. It is still unclear whether the absence of an age gradient in some galaxies supports a different spiral formation scenario, or~it can be explained by some other factors.
\par
Regarding the corotation radius, several methods have been developed to estimate its position based on different assumptions. The~location of the CR can be determined as the radius at which the stellar age gradient changes direction (``offset'' method)~\cite{Oey2003,Tamburro2008,Shabani2018,Abdeen2020,Abdeen2022}. It can also be estimated using the pattern speed measured by the ``T-W'' method~\cite{Rand2004,Hernandez2005,Beckman2018,Williams2021} or from a realistic simulation of a specific galaxy (``model'')~\cite{Garcia-Burillo1998,Rautiainen2008}. In~addition, some authors have considered morphological features predicted to occur near resonances, (``morph'')~\cite{Elmegreen&Elmegreen1995}. The~CR position can be estimated by analyzing the kinematic data of gaseous component, (``F-B'')~\cite{Font2011,Font2014a,Beckman2018}. Analyzing the shift between the potential well and the spiral density can provide a location of corotation resonance (``potential-density'' method)~\cite{Buta&Zhang2009}. In~some studies, this position is determined based on the metallicity distribution~\cite{McCall1982,Scarano2013} or the azimuthal distribution of matter along the spiral arms~\cite{Marchuk2024a}. However, despite the variety of features related to the corotation resonance position, as~shown in our previous paper~\cite{Kostiuk2024} the majority of studied galaxies do not show consistency in CR measurements. This can be interpreted in several ways, ranging from the assumption that some measurements are unreliable to the existence of a significant number of galaxies that do not exhibit a density wave. To~investigate this issue, we have decided to examine some galaxies from  our previous study~\cite{Kostiuk2024} in more detail.
\par
We plan to analyze the positions of corotation resonances measured by different methods in this study and in other studies. By~considering the consistency of these measurements, we will make predictions about the nature of the spiral structure in selected galaxies. We will also investigate other features related to the mechanisms of spiral formation that may support our assumptions. A~similar approach has also been used in several previous studies. The~authors of~\cite{Beckman2018} compared the positions of the corotation radius estimated by ``T-W'' and ``F-B'' methods, and~demonstrated that the spiral pattern in NGC~3344 is related to a mode-coupling process by considering interlocking resonances. In~addition, the~author of~\cite{Marchuk2024c} examined the same features of galaxies with multiple patterns by using CR measurements obtained from a larger number of methods.
\par
Thus, in~this paper, we aim to examine as much observational evidence as possible that supports one or the other theory of spiral formation. In~Section~\ref{sec:measurements} we analyze the accuracy of CR measurements collected for each object from other studies and evaluate the validity of their overall distribution. Then, we obtain independent measurements of the corotation position using the methods described in Section~\ref{sec:method_application}. Furthermore, we compare these measurements with those taken from literature and test separated evidence that both provide a reliable corotation position and contribute to understanding the nature of the spiral structure in studied galaxies (see Section~\ref{sec:results}). Finally, in~Section~\ref{sec:discussion}, we consider our findings in light of the results of other studies and in terms of future research. \mbox{Appendix~\ref{Ap:Vel_maps} and~\ref{Ap:decomp}} provide supplementary information on the application of the methods for corotation radius~estimation.

\section{Objects Under~Investigation}
\label{sec:measurements}
\begin{figure}[H]
\includegraphics[width=13.5 cm]{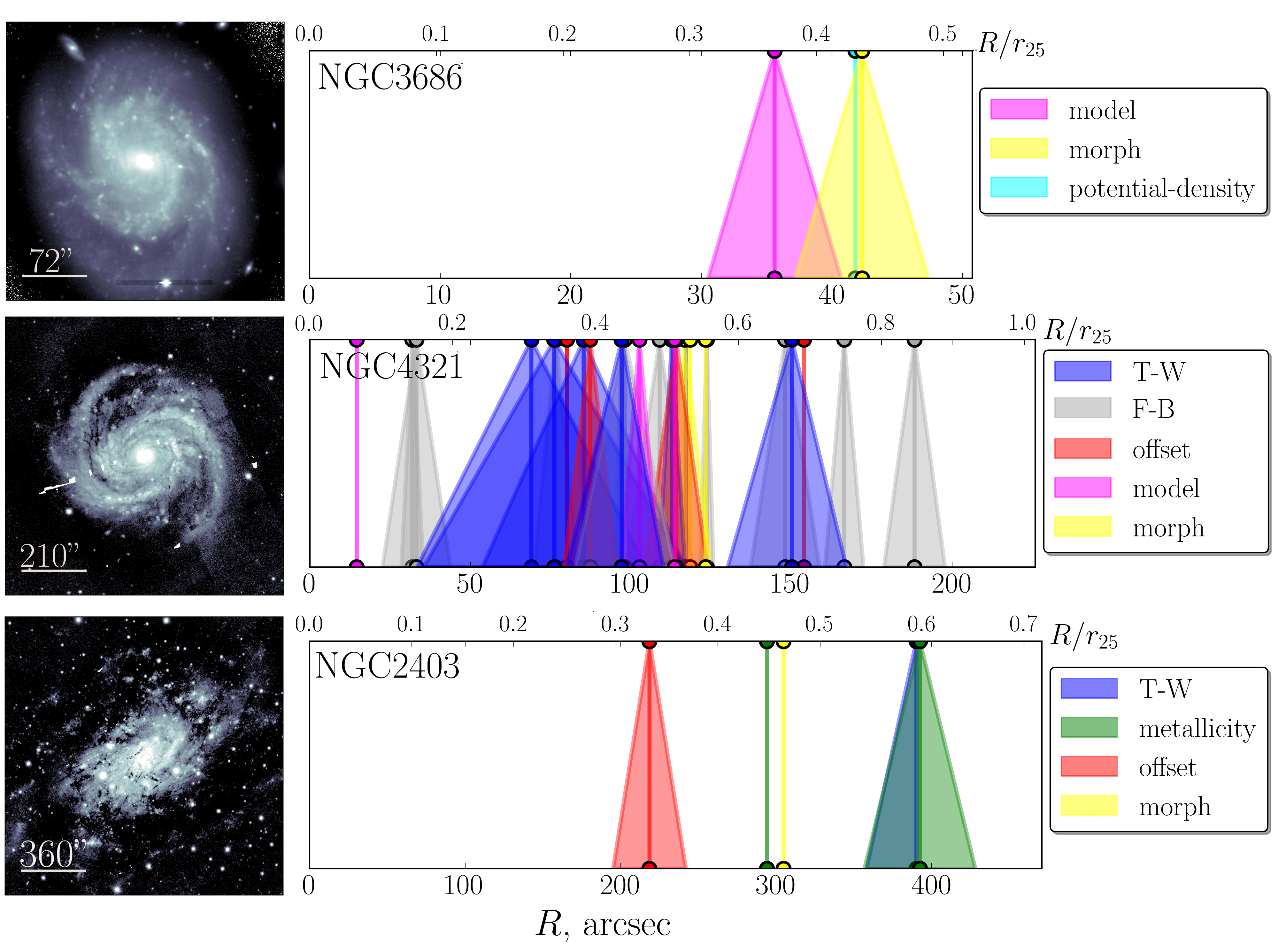}
\caption{\label{fig:distr} The distributions of corotation radius measurements taken from literature are shown on the right. The~shaded areas indicate the error of each measurement (the base of the cone), and~their color indicates the method used (for details and references, see the text). The~lower scale shows corotation positions in~arcsec, while the upper scale shows their positions relative to the optical radius (from NED). To~the left of each distribution, the~image of the corresponding galaxy is shown (NGC~3686 from the Legacy Survey, NGC~2403 and NGC~4321 from PanSTARRS Project).} 
\end{figure}

In this section, we will discuss the basic properties of target galaxies and analyze the corotation radius positions found for these objects in literature. In~our previous study, we collected a dataset of corotation positions obtained using different methods for over 500 galaxies (see~\cite{Kostiuk2024}, for~a more detailed description of the gathered data). The~data is publicly available at GitHub~\endnote{\url{https://github.com/ValerieKostiuk/CRs\_dataset} (accessed on 1 March 2024) 
}. In~this paper, we selected three galaxies that demonstrate a significant difference in the distribution of corotation measurements (see Figure~\ref{fig:distr}). These objects also show a distinct morphology of spiral structure. NGC~4321 is a typical example of a grand-design galaxy, while NGC~3686 has fainter arms than NGC~4321 and the spiral structure of NGC~2403 is flocculent, with~many short patterns. Additionally, these galaxies have markedly different stellar masses
, and~all the data we need for our analysis are publicly~available.
Having a dataset of corotation positions for a particular galaxy, we can predict the nature of the spiral structure in that object. For~instance, the~distributions obtained for NGC~3686 and NGC~2403 have consistency among some measurements, which suggests the presence of one or more corotation resonances in their disks. This implies the presence of a single or several spiral density waves rotating at their angular velocity constant with radius. Meanwhile, the~inconsistencies revealed for NGC~4321 indicate that it is likely that this object has no clear corotation radius, or~that its position may be located almost anywhere in the disk. This case points out to the dynamic spirals, which pattern speed varies with radius as angular velocity of the disk. However, these are only predictions, and~in order to confirm our assumptions, it is first necessary to verify each measurement individually. After~this, we will apply some other methods to obtain independent CR measurements and perform some tests confirming or contradicting one of the~scenarios.

\begin{table}[H] 
\caption{This table presents the parameters of the selected galaxies that were utilized in this~study.\label{tab1}}

\begin{adjustwidth}{-\extralength}{0cm}
\begin{tabularx}{\fulllength}{CCCCCCC}
\toprule
\textbf{Object Name}& \textbf{Stellar Mass, M$_{\odot}$}& \textbf{Absolute Magnitude, Mag}\textsuperscript{1}	&\textbf{Distance, Mpc} &  \textbf{Positional Angle (PA), Deg} & \textbf{Inclination (i), Deg}\textsuperscript{2} & \textbf{Systemic Velocity (V$_{sys}$),~km/s}\textsuperscript{3}\\
\midrule
NGC~3686& 8.3$\times$10$^9$~\cite{Nersesian2019} & $-$18.7 & 21~\cite{Ashby2011}    & 19~\cite{Sorai2019}  & 35 $\pm$ 2~\cite{Sorai2019} &  1152 $\pm$ 1    \\
NGC~4321& 4.9$\times$10$^{10}$~\cite{Nersesian2019}& $-$20.5  & 15.9~\cite{Sheth2010} & 151~\cite{Chemin2006} & 38 $\pm$ 2~\cite{Chemin2006}  &  1561 $\pm$ 2     \\
NGC~2403& 2.9$\times$10$^{9}$~\cite{Nersesian2019} & $-$18.2 & 3.2~\cite{Makarov2014}    & 125~\cite{Hernandez2005} & 60 $\pm$ 2~\cite{Hernandez2005} & 144 $\pm$ 2 \\
\bottomrule
\end{tabularx}
\end{adjustwidth}
\noindent{\footnotesize{\textsuperscript{1} The absolute magnitude values were determined using the apparent magnitudes of the g-band from the SDSS and the distances listed in this~table. \textsuperscript{2} The inclination values presented in this column, which include errors, were used to generate a set of parameters as described in Section~\ref{sec:F-B}. \textsuperscript{3} These values were chosen as they were more suitable for creating a velocity field model in Section~\ref{sec:F-B}. }}

\end{table}
\unskip

\subsection{NGC~3686}
NGC~3686 is a two-armed spiral galaxy with a weak bar. It belongs to the group U~376, without~a clear evidence of interaction~\cite{Marino2013}. A~distribution of corotation radius positions presented in Figure~\ref{fig:distr} (top panel) demonstrates consistency among three independent measurements obtained using different methods.
\par
One measurement was obtained by ``potential-density'' shift method in~\cite{Buta&Zhang2009} (cian lines). According to this paper, the~corotation position is defined as the galactocentric distance at which the phase shift between the potential minimum and the spiral density wave changes from positive to negative. Figure~9 from this paper clearly demonstrates a change in the azimuthal shift from positive to negative at a location of 43~arcsec.
\par
The second CR position (42.3~$\pm$~5.1~arcsec marked by yellow cone) was estimated based on morphological features. According to~\cite{Elmegreen&Elmegreen1995} the position of the outer Lindblad resonance (OLR) is assumed to coincide with the optical radius r$_{25}$ with a precision of 10\%. In~this case, using the rotation curve and the OLR location, the~authors measured the pattern speed and, consequently, the~corotation position with some uncertainty.
\par
The last CR position (36.5~$\pm$~5.1~arcsec indicated by pink cone) available in the collected dataset corresponds to a bar structure, that was found by constructing a galactic model in~\cite{Rautiainen2008}. The~model of the disk in this paper consisted of collisionless stellar particles and inelastic, colliding gaseous particles in a fixed potential. The~pattern speed $\Omega_p$ was a free parameter that was varied to fit a more suitable model with the same morphological features as a real galaxy (see Figure~4 in~\cite{Rautiainen2008}). The~corotation radius was then estimated based on the value of the bar pattern speed that was found. Since this measurement does not associate with spiral pattern, we will not compare it to other corotation positions but keep in mind its location. Besides, authors of~\cite{Rautiainen2008} also estimated the length of the bar $R_{\rm bar} = 23\pm3$~arcsec, which value we will use in this paper further.
\par 
\subsection{NGC~4321 (M~100)}
NGC~4321 is a barred spiral galaxy located in the Virgo cluster. It has two dwarf companion galaxies, and, according to a large HI extension, there may be a tidal interaction with one of these galaxies~\cite{Knapen1993}. There have been numerous studies dedicated to investigating the spiral structure of this galaxy, particularly to locate the position of the corotation resonance of both bar and spiral arms. The~collected dataset consists of 24 values found in 13 research papers, showing measurements at almost each galactocentric radius (as seen in the middle panel of Figure~\ref{fig:distr}). 
\par
The largest number of measurements were found using the ``F-B'' method (gray cones). These positions were determined using velocity fields of ionized (H$\alpha$) and neutral (HI) hydrogen. Among~all the peaks found on ``zero-velocity'' histograms (see Figure~2 from~\cite{Font2011} and Figure~1 from~\cite{Font2014c}), the~authors highlighted two of them, indicating the presence of nested bar. By~comparing their results with those obtained by ``T-W'' in~\cite{Hernandez2005} (see blue cones), the~authors concluded that there is corotation of the inner bar at 20-30~arcsec and the outer one at 90-110~arcsec. As~for other measurements, since the authors did not specify their nature, they will be considered as other non-corotational resonances (i.e., ultraharmonic or Lindblad's resonances). 
\par
 As for the application of the ``T-W'' in~\cite{Hernandez2005}, the~authors divided the galaxy into inner and outer bars and spiral arms, and~based on the H$\alpha$ data, they found the pattern speed and corresponding CR for each of these structures. In~addition to the mentioned corotation position of the double bar, the~CR for the spiral structure was estimated to be approximately 150~arcsec, which also matched the resonance location found using the ``F-B'' method. This approach was also applied by~\cite{Rand2004}. However, despite the convincing regression shown in Figure~16, the~estimated CR position of spirals at 117~arcsec differs significantly from that found in the previously mentioned paper. Since the method initially assumes a constant pattern speed for the spiral structure, the~disagreement between the measurements likely indicates a varying angular velocity with radius of the pattern. Additionally, it should be noted that the pattern speed and corotation position of a bar were also measured recently by the same method in~\cite{Williams2021} with both stellar and gaseous data taken from MUSE and ALMA. Although~these positions are consistent with each other and the CR mentioned earlier, the~range of possible resonance positions is quite large (from 30 to 120~arcsec). Furthermore, based on the figures presented in this paper, it is difficult to assess the quality of the linear regression used to estimate $\Omega_p$ (fig.~29 only shows positive values of velocity and radius, while there should be negative one, and~the point spread does not seem to correspond to the error estimates).
\par
The authors of~\cite{Garcia-Burillo1998} conducted hydrodynamic simulations of a galaxy similar to NGC~4321 and found that the angular velocities of the bar and spiral arms were 160~km/s/kpc and 23~km/s/kpc, respectively. However, the~corotation resonances associated with these pattern speeds (pink cones and vertical lines), at~14 and 102~arcsec, are not consistent with the corotation position identified above. This may be due to the fact that the fitted model does not accurately reflect the morphological features of the galaxy as realistically as other studies using the same method, such as~\cite{Rautiainen2008}. Additionally, several values were measured based on predictions from density wave theory regarding the occurrence of specific morphological features near corotation resonance~\cite{Elmegreen1992,Elmegreen&Elmegreen1995} (see yellow cones), which also did not align with the positions of CR found for spiral arms.
\par
It is also worth noting the measurements obtained using the ``offset'' method~\cite{Oey2003,Abdeen2020} (red cones and lines). According to this approach, the~CR position corresponds to the radius where the age gradient of stars changes direction, providing strong support for the density wave theory. The~authors of these studies traced the spirals formed by different stellar populations and identified the radius at which the patterns change configuration. The~CR location of 147~arcsec found in~\cite{Oey2003} agrees with measurements from the ``F-B'' and ``T-W'' methods. The~authors of~\cite{Abdeen2020} estimated this position using two methods of logarithmic spiral fitting, obtaining two values of 87 and 114~arcsec. These CR positions are not consistent with each other or with other measurements obtained using the same method. They also match the range of corotation radii associated with the bar pattern. This inconsistency in the CR positions can be attributed to both the absence of an expected age gradient and the strong sensitivity of this method to the technique of logarithmic spiral~fitting.
\par
Therefore, despite the fact that most of the analyzed measurements were obtained reliably based on the principles of each method, we cannot highlight the trustworthy CR positions of bar and spiral structures based on the considered distribution. Additionally, we noted inconsistency between CR positions, especially among those obtained using similar approach, such as ``T-W'' and ``offset''. This most likely indicates the absence of specific features predicted by density wave~theory.

\subsection{NGC~2403}
The last object under consideration is a nearby spiral galaxy of SABcd type, which belongs to the M~81 group. It was found some evidence of a possible interaction with a dwarf galaxy companion, DDO~40~\cite{Carlin2019,Veronese2023}. Based on the distribution shown in the bottom panel of Figure~\ref{fig:distr}, this galaxy may have multiple positions of corotation resonances, which suggests the existence of several spiral patterns.
\par
One of the possible CR position located at approximately 300~arcsec were confirmed by two measurements. One of these measurements (green vertical line) was estimated based on an analysis of the oxygen abundance distribution in~\cite{McCall1982}. The~other measurement, marked by yellow vertical line, was based on the assumption that the location of resonances influences the size of the area with prominent spiral and disk structures, as~well as the distribution of H{\sc ii}, as~shown in Figure~7 of~\cite{Roberts1975}. Despite the fact that in both studies, the~measurement errors were not estimated and not so modern observational data were used, these measurements are consistent with each other and confirm the location of the CR in different ways.
\par 
The next two consistent measurements were made using the Tremaine-Weinberg method in~\cite{Fathi2009} (dark blue cone), and~by using the abundance profile in~\cite{Scarano2013} (green cone). In~the first case, a~relatively large error was caused by the sensitivity of the resonance location to uncertainties in the pattern speed. For~the other measurement, obtained using the metallicity gradient method, the~error magnitude is typically associated with the sparseness of the abundance data and makes it difficult to accurately determine the position of CR.
\par
Finally, the~last CR position (see red cone) was estimated using the age gradient method, according to which spirals should exhibit an azimuthal offset between different stellar populations. If~a galaxy has a steady density wave, the~corotation radius should be located at a distance where the profile of the azimuthal offset changes sign. In~\cite{Tamburro2008}, the~authors considered the azimuthal shift between young stellar clusters and molecular clouds traced by 24 $\upmu$m and H{\sc i} images, and~found a sign change at 218~$\pm$~23~arcsec. In~the next section, we will discuss these findings in more detail and show how they can be interpreted in another way.   
\section{Applications of~Methods}
\label{sec:method_application}
\unskip
\subsection{Font-Beckman~Method}
\label{sec:F-B}
As mentioned in~\cite{Kalnajs1978}, the~radial streaming motions of gas are expected to change direction around the resonance. Thus, for~the trailing arms, this motion is towards the center inside the corotation radius and outward outside it. This is the key idea lying at the basement of the variety of methods developed to find the position of resonances in the external galaxies (e.g.,~\cite{Sakhibov89,Canzian1993,Sempere1995,Lyakhovich1997}). In~this paper, we consider one of the approaches described in~\cite{Font2011} and successfully applied to numerous galaxies~\cite{Font2014a}. Unlike the other methods used in the aforementioned papers, this one is based on analyzing residual velocities rather than radial velocities, making it much easier to implement and applicable to multiple objects. The~values of radial velocities depend strongly on the chosen inclination and positional angle of the object under consideration. Therefore, the~use of residual velocities can help avoid uncertainties caused by this effect.
\par
The realization of this method lies on searching the positions of so-called ``zero'' velocities located near the resonances. In~the ideal, noise-free case, these locations indicate the places where streaming motions induced by density wave(s) change their direction and the flow vector is equal to zero. As~mentioned before, ``F-B'' approach uses the residual velocity, the~radially projected component of the departures from systematic galactic rotation, denoted as:
\begin{linenomath}
\begin{equation}
   V_{\rm res} = \dfrac{V_{\rm los}-V_{\rm sys}}{\sin i} - V_{\rm rot} \cos i,
    \label{residauls}
\end{equation}
\end{linenomath}
where $V_{\rm res}$ and $V_{\rm los}$~--- residual and line-on-side velocities, $V_{\rm sys}$ and $V_{\rm rot}$~--- systemic and rotation velocities. In~order to obtain the residual velocity data, we used H${\alpha}$ velocity field taken from Loose groups (NGC~3686)~\cite{Marino2013}, VIRGO (NGC~4321)~\cite{Chemin2006} and SINGS (NGC~2403)~\cite{Daigle2006} surveys using the Fabry-Perot observations. We evaluated the rotation model for each galaxy using the classical 2D tilted-ring modeling of a galaxy with the \textsc{BBarolo} tool~\cite{Teodoro-Fraternali2015}, which allows us to take into account variations in parameters such as position angle, inclination, systemic velocity, and~others. In~this study, we set the parameters in such a way that the program minimizes the residual velocity in each ring by varying the position angle and rotational velocity. Note that the method is quite sensitive to the presence of NaN pixels, which is why we had to limit the radius of the farthest ring. According to the images presented in Appendix~\ref{Ap:Vel_maps}, the~fitted rotational velocity models are in good agreement with the observational velocity fields, which is also confirmed by the small absolute values of the residual velocity distributions.
\par
Based on the technique proposed by~\cite{Font2011}, in~order to estimate the zero velocity position, we considered the radial slit centered on each pixel of the residual velocity map. A~pixel is expected to be at the position of zero velocity if two conditions are satisfied: (1) the two adjacent pixels have different signs of residual velocity, and~(2) the magnitude of the residual velocities of the two adjacent pixels exceeds the uncertainty determined by the spectral resolution of the instrument. The~first condition is associated with changing the direction of gas motions. The~second one is due to noise in the observational data, which can cause a transition from negative to positive velocities, or~vice~versa. 
Next, by~fixing the radii at which zero-velocity points occurred, we created a distribution of these positions. According to~\cite{Font2011}, the~peaks in this histogram correspond to resonance positions.
\par
\begin{figure}[H]
\includegraphics[width=10.5 cm]{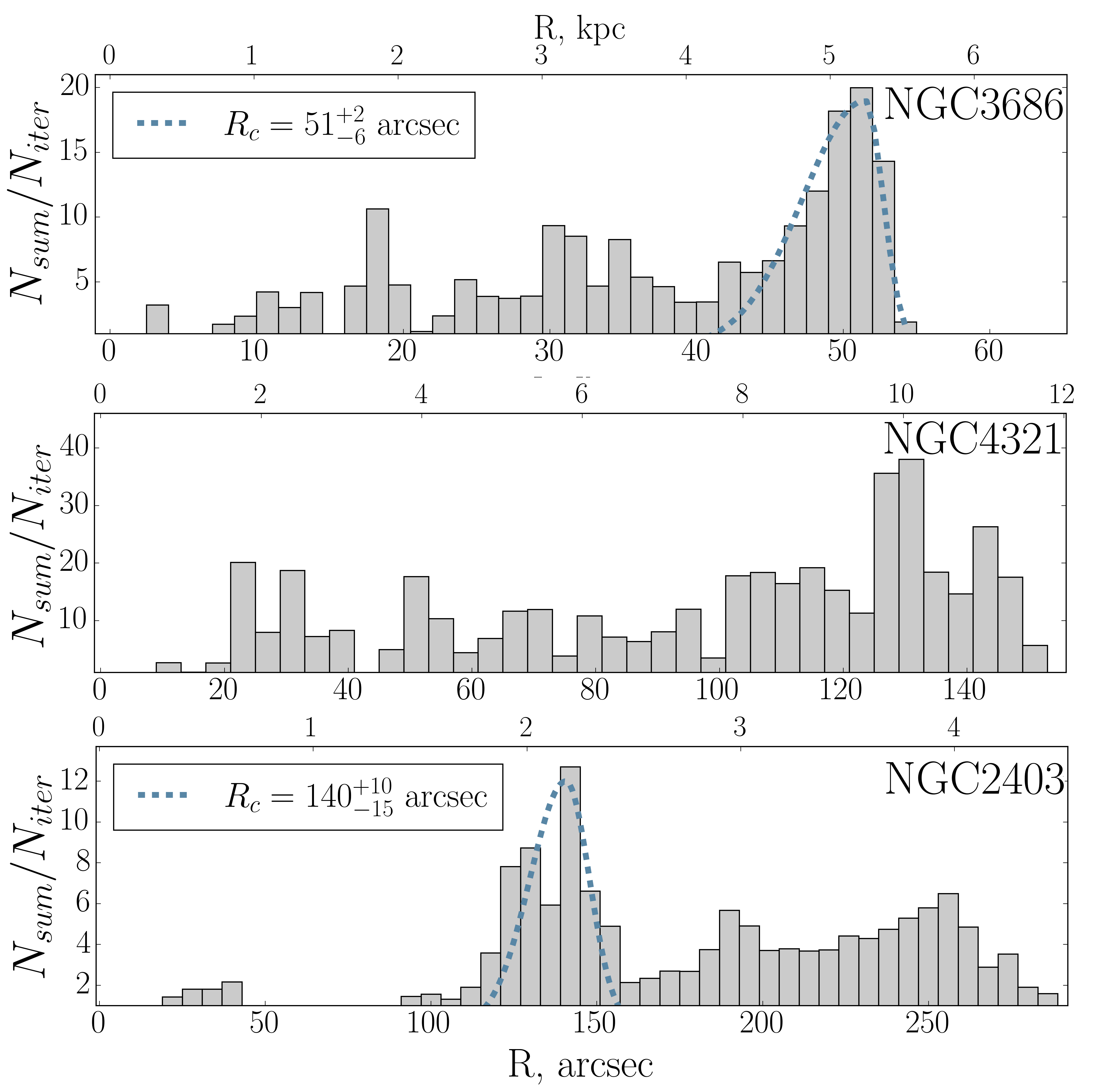}
\caption{\label{fig:F-B} The top, middle, and~bottom panels represent the distributions of average number of zero velocity pixels for NGC~3686, NGC~4321, and~NGC~2403, respectively. The~average was calculated as the sum of distributions from all realizations, included different sets of parameters, normalized to the number of iterations. For~galaxies NGC~3686 and NGC~2403, we identified the most likely corotation position by asymmetric Gaussian function (see the blue dotted lines). The~lower axis is in~arcsec, while the upper ones in~kpc.}
\end{figure}

In this study, we modified the method described above by considering the influence of selected galactic parameters on the resulting distribution of zero-velocity points. For~each object, we generated 600 random samples of these parameters following a Gaussian distribution for inclination ($i$), systemic velocity ($V_{sys}$), and~ring width. The~median values and standard deviations of $i$ and $V_{sys}$ considered for each galaxy are presented in the Table~\ref{tab1}. It should be noted that we also varied the initial approximation for position angle (PA), however, this did not influence the results, as~these parameters were adjusted by a program for each ring. Note that the ring width is 2-4 times greater than the distance between adjacent rings. This ratio depends on the noise level of the initial velocity field data, and~thus it determines how smooth a velocity curve we need to use in order to obtain a more suitable model of rotational velocity. Finally, we obtained 600 distributions of zero-velocity locations for each galaxy, which were then combined. This procedure allowed us to identify the most probable resonance locations and smooth out the noise points. Figure~\ref{fig:F-B} shows the sum of distributions of gas zero-velocities for each galaxy, normalized by the number of iterations ($N_{iter}=600$). It is noteworthy that this modification takes into account not only parameter uncertainties, but~also estimates the corotation range more accurately (see blue dotted lines on the top and bottom panels).
\par 
The histogram obtained for NGC~3686 shows a clear maximum at 50~arcsec, outlined by an asymmetric Gaussian function, to~determine the boundaries of the corotation position. Additionally, there is a minor peak at 20~arcsec that is close to the length of the bar structure determined in~\cite{Rautiainen2008}. In~contrast, the~distribution derived for NGC~4321 shows a somewhat ``fence-like'' pattern with many evident peaks, which may suggest that this technique is sensitive to changes in parameters or/and the occurrence of multiple resonances. The~first explanation may have a basis, as~two variations of this method applied to the same velocity field reveal distributions of zero-velocity peaks with a shift (see Figure~2 from~\cite{Font2011} and Figure~1 from~\cite{Font2014c}). Nevertheless, this shift is not substantial and the histograms remain quite similar to each other and to the one presented in this paper, which confirms the validity of the approach and our modification. Regarding the multiplicity of resonances, some of them could be attributed to inner and outer bar resonances, as~was highlighted in~\cite{Font2011,Font2014c}. The~other peaks outside the bar (60~arcsec) are likely related to spiral patterns, but~it is unclear whether their multiplicity indicates the presence of multiple corotation radii.
The bottom panel of~\ref{fig:F-B} displays the distribution of points with zero velocity for the last considered galaxy NGC~2403. It shows an explicit maximum which is described by a Gaussian function with parameters presented in the legend. Additionally, there is a slight increase in the radius range 200–250~arcsec, which may be more likely to be attributed to noise in the observational data (see Figure~\ref{fig:vel_maps}).  
 
\par
\subsection{Stellar Age~Gradient}
\label{sec:offset}
One of the main observational evidence supporting the spiral density wave theory is based on the investigation of so-called stellar age gradients. In~other words, in~a galactic disk that exhibits a spiral structure, which rotates at a fixed angular velocity, we can see an azimuthal transition of stars of different populations. It is connected to the fact that within the corotation radius, matter in the disk moves faster than the spiral pattern ($\Omega_{\rm disk}>\Omega_{\rm pattern}$), while outside this radius, it moves slowly. There are two different scenarios of age gradients propagation across spiral arms, which do not contradict the density wave theory. According to the first point of view~\cite{Roberts1969}, the~spiral density wave compresses moving with the disc gaseous clouds and induces the star formation process at one edge of the spiral arm. In~the central region, the~disk overtakes the spiral, so star formation occurs on the inner side of the arms. In~the external regions of a galaxy, stars are born on the outer sides of the spiral arms. During~the movement of stars across the spiral, stellar evolution also occurs, so older stars should be located on different sides of the arms relative to younger stars. On~the other hand, the~older population makes up the majority of the total galactic stellar mass and determines its potential. As~a result, this component determines the region where the spiral wave propagates and causes gravitational instability in the gaseous material, leading to star formation. In~this case, the~older stellar population is expected to be located at the inner edge of the spiral arm, inside the corotation radius, and~vice~versa, outside of it. A~clear visual representation of discussed above interpretations can be seen in Figure~2 of~\cite{Miller2019}. It is worth noting that both of these scenarios have been supported by observational evidence (see~\cite{Martinez-Garcia2013,Miller2019,Martinez-Garcia2023}).  
\par
The method of stellar age gradient is based on analyzing the radial variation of the angular offset ($\Delta \phi$) between the spiral patterns traced by stars of different ages (or between stars and H{\sc i} gas). The~radius at which $\Delta\phi$ changes its sign determines the corotation radius location. Note that this sign change occurs regardless of which age gradient scenario is taken place. In~addition to this sign change, there is also strong evidence in support of the density wave theory, which implies that the shape of the angular offset profile is described by a specific function (see~\cite{Tamburro2008}):
\begin{linenomath}
\begin{equation}
   \Delta \phi(r) = (\Omega_{\rm disk}(r)-\Omega_{\rm pattern}) t,
    \label{eq:offset}
\end{equation}
\end{linenomath}
where $\Omega_{\rm disk}$ and $\Omega_{\rm pattern}$ are the angular velocities of the disk and spiral pattern, respectively, and~$t$ represents the time it takes a star to travel from one location to another during the crossing of the spiral.
However, despite the apparent simplicity of the problem, there is disagreement about the existence of a stellar age offset in real galaxies. According to theoretical predictions~\cite{Roberts1969} and hydrodynamic simulations of galaxies~\cite{Dobbs2010,Pettitt2017}, if~a global, steady density wave exists, then an angular offset is also expected. However, from~an observational point of view, there seems to be a lack of consensus. Some authors have found a clear angular offset~\cite{Tamburro2008,Egusa2009,Yu&Ho2018,Abdeen2022}, while others have not found it in observed galaxies~\cite{Foyle2011,Martinez-Garcia2013,Choi2015,Ferreras2012}. It is also worth noting that, among~a series of studies examining this issue for a particular galaxy, one half found an age gradient while the other half did not~\cite{Vallee2020}. Such disagreement could be caused by differences in methodology used to verify the expected offset or by differences in tracers~\cite{Louie2013}. In~one series of papers, authors analyze cross-correlation between rings cut from images to find angular offsets~\cite{Tamburro2008,Egusa2009,Foyle2011,Egusa2017}, while others measured distances between stellar clusters~\cite{Sakhibov2021}. Some studies used spatial isochrones~\cite{Oey2003,Abdeen2022}, others traced the spiral pattern in images at different wavelengths and compared the pitch angles~\cite{Yu&Ho2018,Abdeen2020,Marchuk2024b}. Additionally, the~ambiguity in the presence or absence of an age gradient can be related to the use of broadband colors to determine stellar ages, due to the age-metallicity degeneracy and reddening effects that are not taken into account in these cases. On~the other hand, there is also a discrepancy in the determination of corotation radii among those obtained not only by the ``offset'' method, but~by many other methods as well (see~\cite{Kostiuk2024}). Therefore, it is possible that for some galaxies, there could be no localized corotation resonance of the spiral structure at all.
\par
In this study, we investigated the spatial offset between the modelled spirals obtained using a decomposition procedure with spiral structure, based on images of galaxies in the ultraviolet and near-infrared ($3.6~\upmu \rm m$). It should be noted that the images in the far-ultraviolet (FUV) and near-ultraviolet (NUV) wavelengths trace the non-ionized, young stellar population, with~ages of approximately 100 and 200 Myr, respectively. Meanwhile, the~near-infrared images indicate the presence of stars with late spectral types, whose ages can reach up to 10 billion years. A~more detailed description of the decomposition process and the data used is presented in the Appendix~\ref{Ap:decomp}. Such procedure of decomposition with spiral arms has proven itself well (see~\cite{Chugunov2024,Marchuk2024b}). Note that this approach allows us to create a more realistic model of the spiral structure compared to studies that use Fourier transformation and logarithmic function to construct spirals~\cite{Davis2012,Pour-Imani2016,Abdeen2020,Abdeen2022}. In~addition, this technique allows us to obtain a smoother distribution of material in the spiral arms, which makes it easier to identify individual arms and detect any offset effects between spirals containing different stellar populations. However, despite these advantages, this method only allows us to analyze the angular shift between the ridges of each spiral. This gives us not a precise offset, but~rather an average value at each radius that depends on the chosen spiral model.
\par
\begin{figure}[H]
\includegraphics[width=11.5 cm]{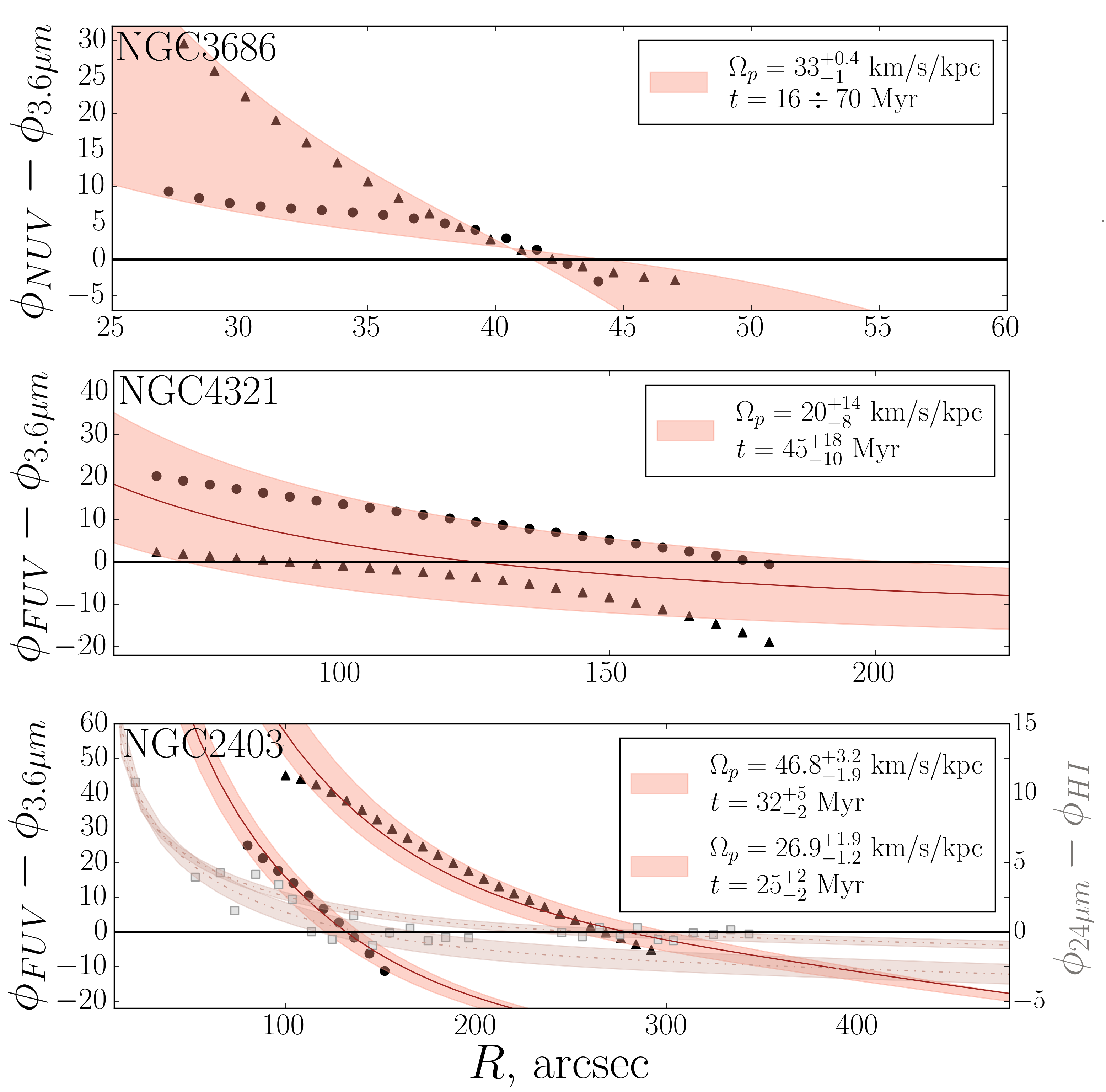}
\caption{\label{fig:offset} Black circles and triangles represent the profiles of the angular offset between the centers of the modelled spirals in the ultraviolet and infrared images of three investigated galaxies (NGC~3686~---top, NGC~4321~---middle, and~NGC~2403~---bottom panel). Each marker corresponds to a different spiral used in the analysis. The~same markers are used to denote patterns in the figures in the Appendix~\ref{Ap:decomp} (middle column). The~red filled area contains a set of curves obtained from Equation~(\ref{eq:offset}), with~a range of pattern speeds and time values presented in the legend. Note that the angular velocity of the disk used in Equation~(\ref{eq:offset}) was calculated for each object based on the rotation velocity profile found in Section~\ref{sec:F-B}. The~gray unfilled squares in the bottom panel are from~\cite{Tamburro2008}, and~their range of values is on the right~\emph{y}-axis.}
\end{figure}

 
Therefore, we obtained the angular offset between the spirals traced in the images of the target galaxies in the far ultraviolet (NUV for NGC~3686) and near-infrared bands, and~we considered their variation with galactocentric radius (see Figure~\ref{fig:offset}, black circles). Red filled areas represent regions with theoretical angular offset profiles (eq.~\ref{eq:offset}), with~the parameters $\Omega_{\rm pattern}$ and $t$ shown in the legend for each plot. The~angular shift profile for NGC~3686 shows a clear sign change at 42-44~arcsec, which corresponds to a pattern speed of 33~km/s/kpc. In~addition, taking into account the angular offset for both spirals, their profile can be described by a series of theoretical curves with a wide range of times. This difference in times can be explained by the wide range of tangential velocities of the stars crossing the spiral arms and the use of model spirals that give not so accurate values of $\Delta\phi$.
Furthermore, it should be noted that a particular star does not necessarily reach the age characteristic of late spectral type stars during the movement from one location to another. Instead, a~specific star may reach the area exhibiting by older stellar populations, enter a potential well, and~continue to evolve.
\par
The angular offset profile for NGC~4321, shown in the middle panel of Figure~\ref{fig:offset}, shows two zero crossings if we consider the two spirals separately. This may suggest the existence of two spiral arms rotating at significantly different angular speeds (34~km/s/kpc and 12~km/s/kpc), which are also located at the same galactocentric distance. This interpretation is difficult to confirm, especially given the symmetrization of these arms. Instead, it would be more appropriate to describe these profiles using a series of curves with a wide range of pattern speeds and time scales. In~addition, angular shift profiles show a gradual decrease, and~considering that its value is not precise, we may assume an almost constant offset $\Delta \phi$ over the radius. This suggests either the absence of a corotation radius as such or its presence over a fairly wide range.   
\par 
Finally, for~NGC~2403, the~bottom panel shows a clear offset between the spirals containing young and old stars for both arms. This galaxy has a complex spiral structure, making it difficult to create a suitable model that accurately represents the shapes of the spirals compared to a real galaxy. For~further investigation, we selected two spirals (hereafter referred to as the inner and outer arms and marked by circles and triangles on Figure~\ref{fig:decomp_models} in the bottom panel) that, in~our opinion, correspond more closely to the real structures. For~the inner spiral, $\Delta \phi$ changes sign at 140~arcsec, while for the outer spiral, it changes at 280~arcsec. This indicates the existence of two spiral structures that are rotating at angular velocities of 47~km/s/kpc and 27~km/s/kpc, respectively. Note that a similar analysis was conducted in~\cite{Tamburro2008}, the~authors used the cross-correlation method to detect the offset between molecular clouds and young dusted stellar clusters ($24~\upmu \rm m$). Their measurements are shown in the same figure as gray squares, and~the range of values is labeled on the right \emph{y}-axis. The~angular offset obtain in that study can also be described by two theoretical curves, rather than a single line that crosses the zero line at 200~arcsec, as~was done in~\cite{Tamburro2008} (see Figure~15). Hence, taking into account the possibility of multiple spiral structures in this galaxy, we can conclude that the results from both methods are consistent and indicate two corotation radius positions.
\par
Hereby, we applied the stellar age gradient method to a sample of three galaxies. Analysis of the angular shift profiles showed that, when considering each spiral arm individually, there is a clear offset between regions with different stellar populations, as~expected in the density wave scenario. However, as~shown for galaxies such as NGC~4321 and NGC~2403, this visual offset could also be interpreted as evidence for dynamic spirals or multiple patterns, rather than a single spiral density wave. Thus, this arisen disagreement about the existence of an age gradient may be caused not only by differences in the methodology used to measure it, but~also by the presence of galaxies with spirals of different~nature.

\unskip 
\subsection{Metallicity~Gradient}
\label{sec:metal}
The corotation resonance is believed to play a significant role in the chemical evolution of galaxies, particularly in the shape of the abundance distribution as a function of galactocentric radius. It could, for~example, be responsible for the sharp drop in metallicity between the inner, steep profile and the outer, flat profile. This ``break'' in the metallicity profile has been observed not only in real data from the Milky Way~\cite{Twarog1997, Lepine2011} and nearby galaxies~\cite{Vila-Costas1992, Martin1995, Scarano2013, Belfiore2017, Sanchez-Menguiano2018, Chen2023}, but~also in hydrodynamical simulations~\cite{Lepine2001, Garcia2023, Hemler2021, Acharyya2024}. This effect is linked to the previously discussed phenomenon of changing directions of gas flows crossing the corotation circle, which leads to bimodality in changes in angular momentum~\cite{Minchev2011}. A~corotation resonance acts as a barrier that prevents the mixing of gas between the inner and outer parts of the disk. As~a result, the~inner disk continues to accumulate metals due to local star formation, while the outer disk remains metal-poor due to its lack of star~formation.

In addition, some recent chemical models (for example,~\cite{Spitoni2019,Spitoni2023}) have shown that significant variations in abundances occur at resonances, particularly at corotation. This is explained by the faster chemical evolution in the vicinity of corotation due to the lack of relative gas-spiral motions. It causes gas overdensities and higher star formation rate (SFR), making metal mixing more efficient.
\par 
It is worth noting that there are other possible explanations for the observed break in the metallicity profile. Some studies, such as those by~\cite{Friedli1994,Roy1997,Friedli1999} associate these breaks with indicators of bar formation. Other authors, such as~\cite{Belfiore2017,Belfiore2018,Simons2021} suggest that the appearance of the metallicity profile is related to gas density, stellar density, and~SFR distribution. There are also several papers that suggest the flattening (or even a positive gradient) in the outer disk's metallicity profile may be due to accretion of pre-enriched material~\cite{Kewley2010,Miralles-Caballero2014,GusevDodin2021}. Some authors argue that spatial resolution may also contribute to the ``flattening'' of observed metallicity profiles~\cite{Carton2017,Acharyya2020}. Therefore, considering the above-mentioned facts, we will use this method with caution. If~the corotation resonance obtained using this method is consistent with other measurements, we will consider it as a slight argument. Otherwise, it will mean that a certain break have another nature.
\par
\begin{figure}[H]
\includegraphics[width=12.5 cm]{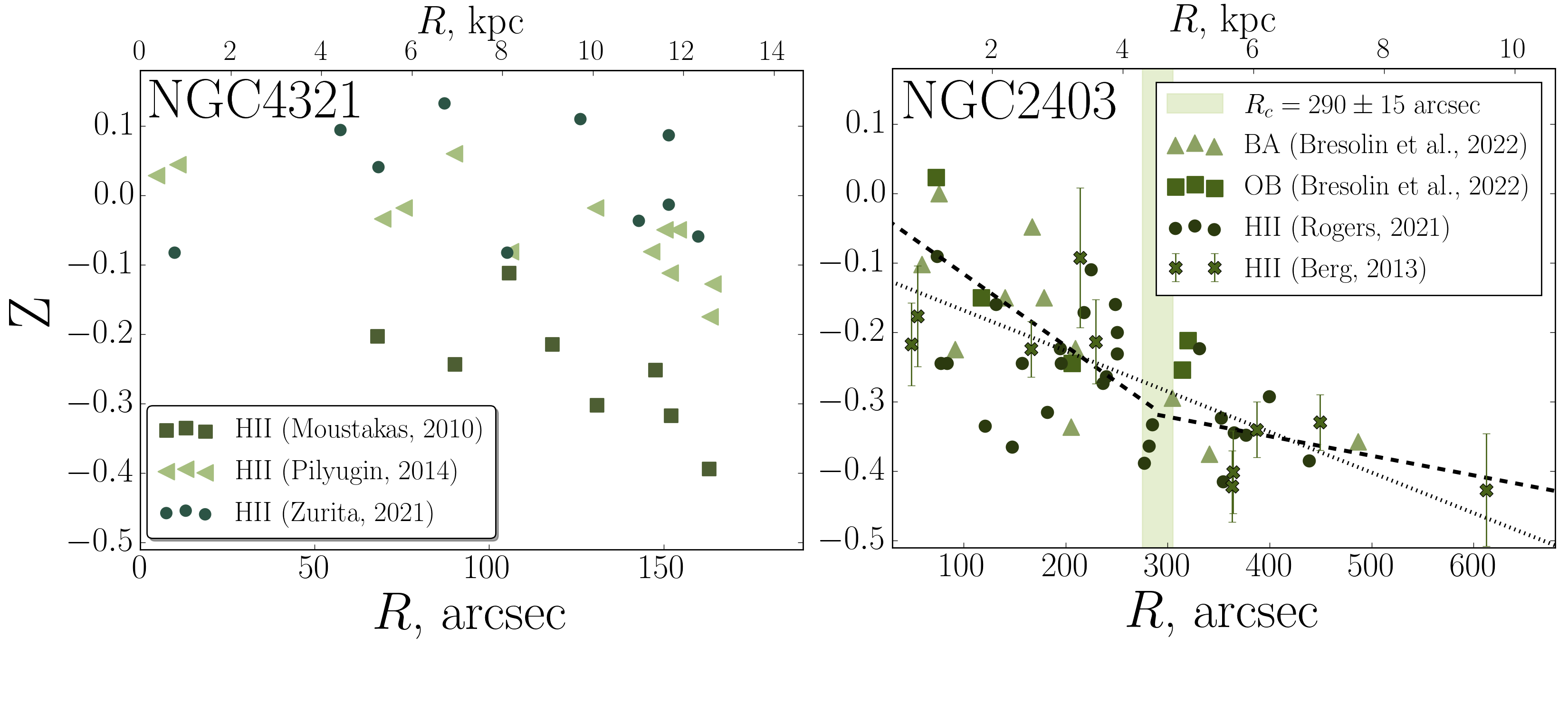}
\caption{\label{metallicity} The galactocentric metallicity radial gradient for NGC~4321 (left) and NGC~2403 (right). All measurements, besides~ones taken from~\cite{Rogers2021} and~\cite{Bresolin2022}, were transformed from oxygen abundances by adopting the solar oxygen abundance value $(12 + \log{\rm O/H})_{\odot}= 8.69$. }
\end{figure}

To apply this method, we collected data on oxygen abundance from various studies, $\log(\rm O/H)+12$, and~analyzed its distribution as a function of galactic radius. Specifically, for~NGC~2403, we found metallicity measurements in~\cite{Berg2013}, obtained for several H{\sc ii} regions (see Figure~2). Based on Figure~5 from that paper, there is a slight drop in metallicity at 4-5 kpc, suggesting the presence of a corotation point there. To~verify our assumption, we also used other measurements such as metal abundances of supergiant B and A stars from~\cite{Bresolin2022}, and~that of H{\sc ii} regions from~\cite{Rogers2021}, the~radial profiles or these data are shown in Figure~\ref{metallicity} (right). Note that both sets of measurements were taken from~\cite{Bresolin2022}, where these measurements have been converted to a common calibration, which allow us to compare them. The~figure reveals a clear change in gradient at 275-305~arcsec (green shaded area), suggesting a possible corotation resonance.
\par
As for NGC~4321, we used abundance data from H{\sc ii} regions measured in~\cite{Moustakas2010, Pilyugin2014, Zurita2021} indicated by squares, triangles, and~circles in the left panel of Figure~\ref{metallicity}. It is not difficult to notice a significant difference between these measurements, which is more likely due to the different calibrations used in those studies than to scattering in abundance.Note that the authors of~\cite{Zurita2021} obtained their measurements by H{\sc ii}-CH{\sc i}-mistry code used the results of~\cite{McCall1985,Shields1991}. Thus, despite the discrepancies between abundances from different studies, none of them show a clear break. Additionally, compared to the metallicity profile of NGC~2403, all sets of measurements show a flatter or even increasing gradient. This indicates efficient mixing processes throughout the disk and implies the absence of a corotation radius. Note that we were unable to perform a similar analysis for NGC 3686, as~there was a lack of necessary data in the literature.
\par
We would like to reiterate that these findings do not necessarily indicate the existence or absence of corotation radii in any particular galaxy. Rather, they should be seen as additional indirect evidence, which should only be considered if there are other strong facts in favor of one or another scenario of spiral~formation.

\unskip

\section{Results}
\label{sec:results}
In this section, we will gather all the estimates of the corotation radius obtained in this study (see Table~\ref{tab2}) and from other papers, and~analyze their consistency. We believe that the existence of a well-defined corotation resonance implies that a galaxy exhibits a steady spiral density wave. A~reliable estimation of the CR position is possible when measurements obtained using different methods are consistent within error limits. Conversely, an~absolute disagreement in measurements may be explained by the abundance of corotation positions, which occurs when the pattern speed coincides with the angular velocity of the disk at almost all galactocentric radii. This last case is likely consistent with the dynamic spirals'~scenario. 

\begin{table}[H] 
\caption{This table presents the corotation radius measurement found in this~paper.\label{tab2}}
\begin{tabularx}{\textwidth}{CCCC}
\toprule
\textbf{Object Name}& \textbf{Font-Beckman Method} & \textbf{Stellar Age Gradient}& \textbf{Metallicity Gradient} \\
\midrule
NGC~3686    & 51$_{-6}^{+2}$~arcsec  & 43~$\pm$~1~arcsec  &     \\
NGC~2403    & 140$_{-15}^{+10}$~arcsec & 140~$\pm$~10, 275~$\pm$~25~arcsec  & 290~$\pm$~15~arcsec \\
\bottomrule
\end{tabularx}
\end{table}
\par
\begin{figure}[H]

\includegraphics[width=12.5cm]{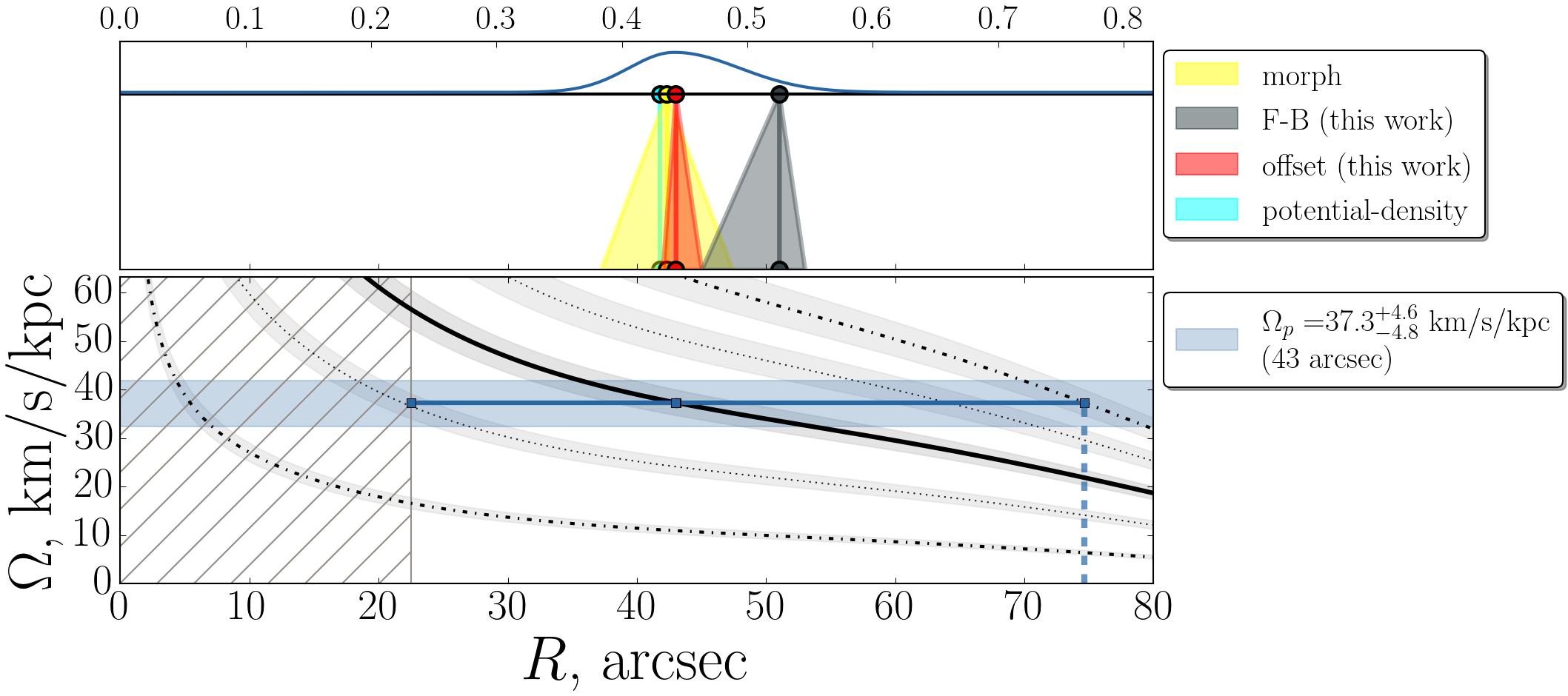}
\caption{\textit{The top panel:} the same as on Figure~\ref{fig:distr} for NGC~3686 including measurement both obtained in this work and found in literature. The~blue line at the top shows the fitted asymmetric Gaussian function. \textit{The bottom panel:} black solid line shows angular velocity profile for the same galaxy. To~the left and right of it, dashed lines define the areas of inner and outer Lindblad resonances. Ultraharmonic resonances are plotted by dotted lines. The~blue horizontal lines connect the positions of the ultraharmonic resonance $\Omega - \kappa/4$, the~corotation radius and the OLR for the angular pattern speed, the~value of which is presented in the legend. The~gray hatched area defines the bar~position. \label{3686_consistency}}
\end{figure}

The distribution of CR measurements for NGC~3686 is presented in Figure~\ref{3686_consistency}, which contains both measurements found in the literature and those obtained in this study. It shows an agreement between the four corotation radii estimated using different methods. To~find the most likely position of CR, we fit the distribution of measurements with an asymmetric Gaussian function (shown by the blue line at the top). We estimate the range of CR to be between 38 and 48~arcsec, corresponding to a pattern speed of $\Omega_p = 37.3^{+4.6}_{-4.8}$~km/s/kpc. The~angular velocity profile (shown as the black solid line) was derived from the velocity curve provided by~\cite{Marino2013}. Errors in the pattern speed take into account uncertainties in the CR position and uncertainties in the velocity curve. In~the case of NGC~3686, the~consistency of CR measurements not only confirms the position of this resonance but also support the existence of spiral density wave in the disk. The~dotted lines in the lower panel of Figure~\ref{3686_consistency} show the profiles of the ultraharmonic resonances at $\Omega\pm\kappa/4$\textbf. The~location of the inner ultraharmonic resonance, which corresponds to the spiral pattern rotating at an angular speed found, is in a range between 18 and 30~arcsec.

Interestingly, the~location of the bar ends, marked by the hatched area, matches the position of this resonance. According to theoretical predictions, spiral density waves propagate within the inner and outer Lindblad resonances, reflecting and amplifying at them and corotation radius~\cite{Goldreich1978,Bertin1989}. Therefore,  we can consider the inner limit to be at the ultraharmonic resonance, which coincides with the bar ends and the beginning of the spiral arms. This finding is not entirely consistent with the conclusions of~\cite{Rautiainen1999}, who suggested that there is likely to be an overlap between the CR of the bar and the ultraharmonic resonance of the spiral pattern. The~position of the bar corotation of 35~$\pm$~5~arcsec is indeed in disagreement with the range of the $\Omega-\kappa/4$ resonance. This could be explained by the inaccuracy of the value found in~\cite{Rautiainen2008}, which also has not been confirmed using other methods. Besides, authors of~\cite{Buta&Zhang2009} using the ``Potential-Density Phase-Shift'' method noticed that their measurements of CR of the bar are systematically lower than those obtained in~\cite{Rautiainen2008}. Moreover, the~position at 27~arcsec where, according to~\cite{Buta&Zhang2009}, is going the decoupling process between the bar and spiral arms in this galaxy occurs (see Figure~9), is consistent with the inner ultraharmonic resonance range.    
Thus, all these arguments, taken together with the localization of the corotation radius obtained by different methods, support the presence of a spiral density~wave.
\par
The analysis performed for NGC~4321 has produced some unexpected results for an object with a clearly visible grand design structure. The~result of the Font-Beckman method suggests that the disk of this galaxy contains many resonances, that could attribute to both corotation and others (see Figure~\ref{fig:F-B}, middle panel). It is worth noting that the authors of~\cite{Font2014c} or~\cite{Font2011} derived almost similar histograms, also highlighted the presence of multiple resonances. Furthermore, we explored the issue of whether or not there is a gradient in the age of stars across the spiral pattern. We discovered that although there is a slight angular offset between the model spirals traced in NIR and FUV images, the~angular shift profile does not clearly cross the zero line at a particular radius (see Figure~\ref{fig:offset}, middle panel). The~range of possible crossings of the CR position, related to these shifts, varies within rough limits of 70 and 200~arcsec. According to the corotation radius definition, this indicates that the behavior of the angular speed of the spirals is more similar to that of a disk than to a constant value through the radius.
\par
\begin{figure}[H]

\includegraphics[width=12.5cm]{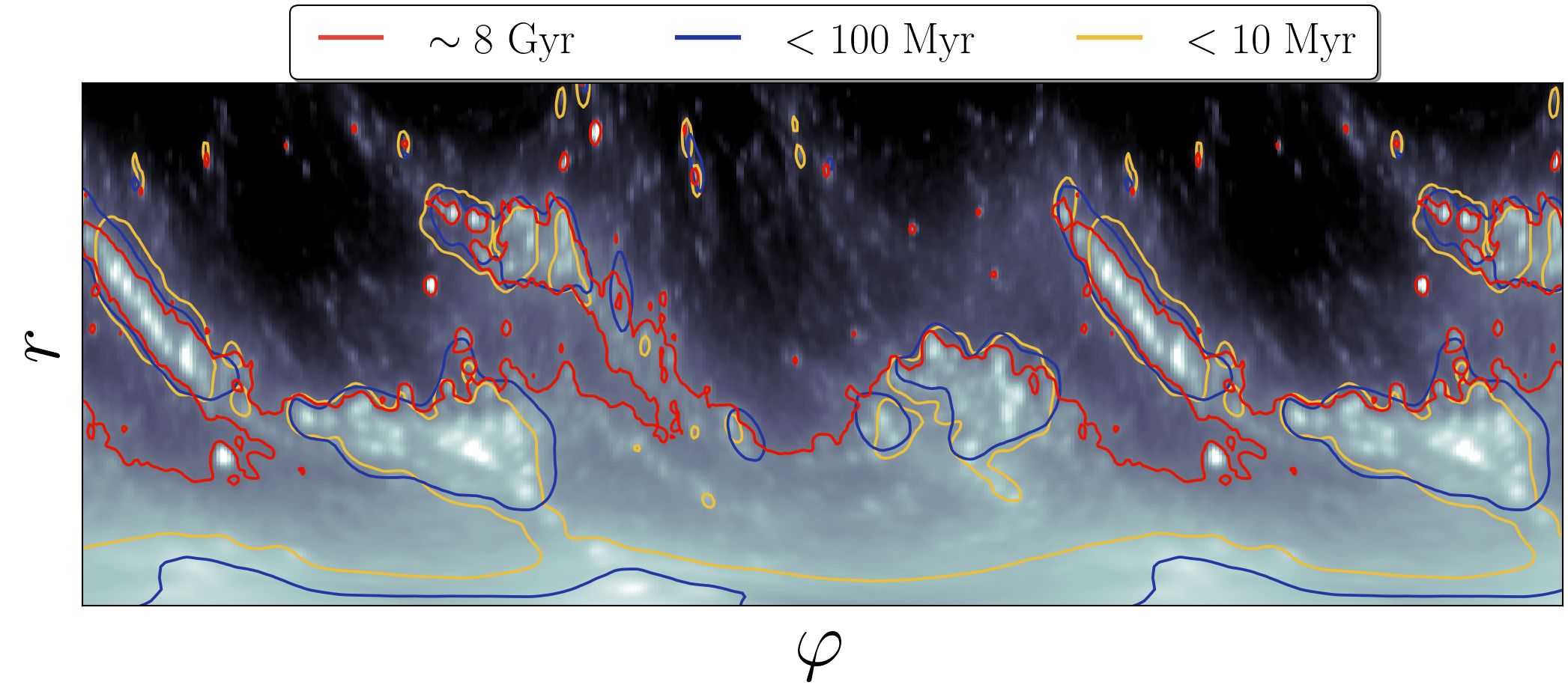}
\caption{Model of the old stellar population for NGC~4321 in polar coordinates. The~horizontal axis represents the azimuth angle, which ranges from -35$^\circ$ to 395$^\circ$, with~the polar angle calculated from the semi-major axis of the galaxy in a counterclockwise direction. The~vertical axis shows distances up to 200~arcsec. The~colored lines represent isolines of equal brightness, which outline spirals traces by three stellar populations (as indicated in the legend). \label{NGC4321_age_gradient}}
\end{figure}

As mentioned earlier, the~direct investigation of the age gradient through color gradients can strongly affect conclusions about the expected offset, primarily due to the age-metallicity degeneracy. To~confirm or disprove the results obtained for NGC~4321, we also analyzed the angular offset of the spirals traced by different stellar populations using images taken from~\cite{Nersesian2020}. In~this paper, the~authors obtained 2D maps of various stellar components (see Figure~1 of the corresponding paper), each associated with SED template and 3D spatial geometry (for details see~\cite{Verstocken2020}). To~verify the existence of an angular offset across the spiral arms, we used maps of old ($\sim$ 8 Gyr), young, non-ionizing (< 100 Myr), and~ionizing (< 10 Myr) stellar population, deprojected these maps into a galactic plane and transformed them into polar coordinates. As~a result, we compared the mutual locations of the spirals composed of stars of different ages (see the outlined spirals in Figure~\ref{NGC4321_age_gradient}). We found, similar to the previous approach, a~slight angular shift, but~no sign of a changing direction in the age gradient. Additionally, similar behavior in the age gradient across spiral arms was revealed by the authors of~\cite{Sanchez-Gil2011} (see Figure~3 in their paper). Based on this analysis, it can be concluded that NGC~4321 does not exhibit an explicit stellar age gradient that clearly changes direction at the corotation radius. In~addition, we have demonstrated an almost equal distribution of CR measurements obtained using different methods, which apparently implies a varying pattern speed profile rather than a constant one. All these findings suggest that the spiral arms of NGC~4321 are more likely to be formed through a different mechanism than the density wave~theory. 
\par
\begin{figure}[H]

\includegraphics[width=12.5cm]{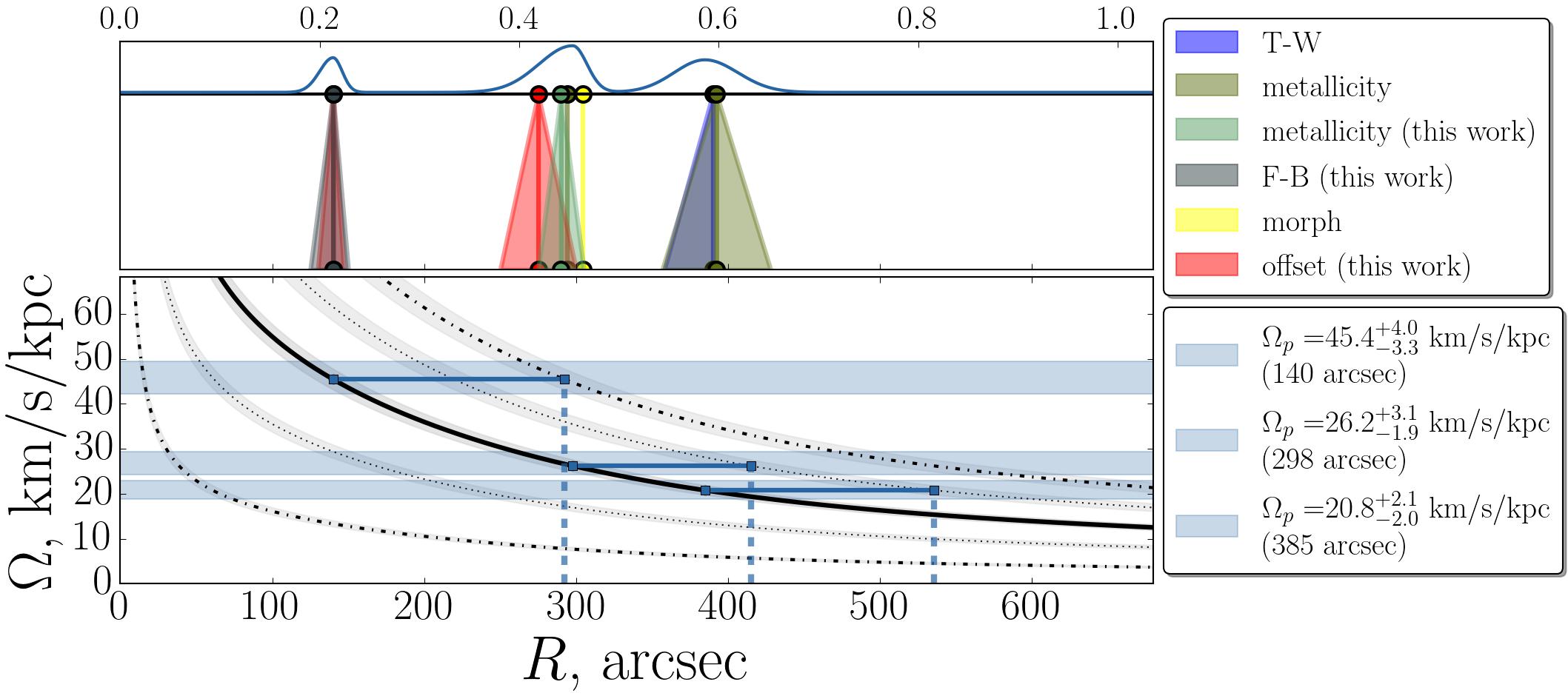}
\caption{\textit{The top panel:} the same as on Figure~\ref{fig:distr} for NGC~2403, including measurement both obtained in this work and found in literature. The~blue line at the top shows the fitted asymmetric Gaussian function. \textit{The bottom panel:} black solid line shows angular velocity profile for the same galaxy. To~the left and right of it, dashed lines define locations of inner and outer Lindblad resonances. Ultraharmonic resonances are plotted by dotted lines. The~blue horizontal lines connect the positions of the corotation radius and the OLR (or ultraharmonic resonance) for the angular pattern speed, the~value of which is presented in the~legend. \label{2403_consistency}}
\end{figure}

As for NGC~2403, the~distribution of corotation resonances shown in Figure~\ref{2403_consistency} demonstrates the consistency of two measurements at approximately 140~arcsec and four ones at approximately 280~arcsec. This may correspond to two localized corotation positions, implying the existence of the same number of patterns rotating at different angular velocities. Moreover, as~shown in Figure~\ref{fig:offset}, the~stellar age gradient profile agrees well with the shape of the velocity rotation profile, and~it clearly crosses the zero line at the vicinity of considered resonances. This provides independent confirmation of their locations. The~bottom panel shows the profile of the angular velocity of the disk by solid line, which was obtained using the velocity rotation curve presented in~\cite{Ponomareva2016}, the~dashed and dotted lines indicate the 2$\div$1 and 4$\div$1 resonances, respectively. Note that the location of the OLR of the inner structure at 300~arcsec coincides with the position of corotation related to the second pattern. There is also the possibility of a third corotation radius at 400~arcsec, which location in turn matches with ultraharmonic resonance $\Omega+\kappa/4$ of the second pattern within error limits. Such overlapping is expected according to theoretical predictions in the case of multiple spiral patterns, which provide an effective transfer of energy and angular momentum outward of the disk~\cite{Masset1997,Rautiainen1999}. This is independent evidence confirming both the existence of CRs and the presence of multiple structures. Note that such mode coupling was previously detected for this galaxy in~\cite{Marchuk2024c}. The~author of~\cite{Marchuk2024c} has also demonstrated the existence of three corotation radii using, among~other measurements, one at 200 arcsec taken from~\cite{Tamburro2008}, which was not confirmed by other methods. However, in~this paper, during~the analysis of age gradient profiles, we have clarified and replaced this position to 140 arcsec, which has also been confirmed by  the ``F-B'' method. Therefore, the~findings of~\cite{Marchuk2024c} differ from our study by the corotation resonance position of the inner pattern and its angular velocity. 
\par
In this section, we considered the CR distributions containing the positions taken from the literature and found in this paper. In~addition, we analyzed the features that support or contradict the density wave scenario, such as the stellar age gradient profile and the interlocking of resonances for each galaxy. Using this approach, we demonstrated that NGC~3686 likely exhibits a density wave, while NGC~4321 may have dynamic spirals, and~the spiral structure of NGC~2403 most likely consists of three different spiral modes. Therefore, starting from a prior prediction of the nature of spirals based on the distribution of CR measurements, we could confirm our initial assumptions discussed in Section~\ref{sec:measurements} using independent observational~evidence.
\unskip

\section{Discussion and~Conclusions}
\label{sec:discussion}
Every single spiral galaxy has a unique spiral pattern that allows not only experts but anyone to distinguish galaxies from each other. There is some bias in our understanding of how galaxies should look if we assume one or another mechanism responsible for spiral structure formation. That way, one undoubtedly refers a grand-design galaxy to that having a steady spiral density wave and a flocculent one to that with transient or recurrent patterns. This point of view had risen from studies of~\cite{Lin1967,Thomasson1990} who demonstrated that spiral patterns with spiral numbers greater than two or three are likely unstable. Flocculent galaxies are believed to form due to local instabilities and perturbations attributed to transient or recurrent spiral structure. In~addition, there has been an assumption that the majority of grand-design galaxies contain a bar or companion structure, which induces a two-armed structure~\cite{Kormendy1979,Elmegreen1982}. However, according to recent studies of~\cite{Hart2016,Smith2022a} it has been shown that there is no significant difference in the number of grand design galaxies found in clusters or in isolation. The~authors of~\cite{Savchenko2020} also noted that flocculent galaxies are common among isolated and barred galaxies. Additionally, some unbarred, isolated galaxies with prominent arms have been observed (for example, see~\cite{Smith2022a}), but~their existence needs an explanation. Also in~\cite{Kostiuk2024} in was noted that the consistency of corotation radius measurements, implying the existence of spiral density wave, does not correlate with the existence of a bar or a specific type of spiral structure (flocculent or grand-design) in a galaxy. Moreover,~\cite{Chen2024} investigated two galaxies with similar morphological features and found evidence that they may have different origins of their spiral arms. All these facts suggest that the type of the spiral structure is not necessarily determined by its nature.
\par
We conducted a comprehensive analysis of the nature of the spiral arms in three morphologically different galaxies. First, in~Section~\ref{sec:measurements}, we gathered and verified measurements of the corotation radius obtained by different methods for each object, and~then based on the distributions of these measurements, we made predictions about the likely positions of CRs and the nature of each galaxy's spiral structure. Furthermore, we performed our own measurements (see Section~\ref{sec:method_application}) to clarify and confirm corotation radius locations. During~the analysis, we also tested for the presence of a stellar age gradient and consistency of the shape of the angular shift profile with that expected in the density wave scenario in Section~\ref{sec:offset}. In~addition, we considered the possible connection between the location of CR and the radial distribution of metallicity (see Section~\ref{sec:metal}). To~further confirm the corotation positions, we investigated the expected resonance overlap in galaxies with bar or multiple patterns. Thus, applying this approach to three chosen galaxies, we obtain quite interesting results, which demonstrate that the relation between the nature of spiral arms and their appearance is quite ambiguous.
\par
Firstly, we would like to draw special attention to the NGC~3686 case, for~which convincing arguments in support of the spiral density wave scenario have been found. Certainly, this is not the first time a spiral density wave has been identified in real galaxies (see~\cite{Pour-Imani2016,Miller2019,Peterken2019,Abdeen2022}). The~study of~\cite{Peterken2019} derived the pattern speed for the galaxy UGC~3825 by using the offset between stellar populations with known ages. As~a direct evidence of the density wave scenario they found an almost constant profile for the angular velocity of the spiral arms. However, according to Figure~2 from this paper, the~variation in $\Omega_p$ profile seems to be comparable to the changes in the angular velocity of the disk. The~authors of other mentioned papers measured the pitch angles of logarithmic spirals describing patterns of different colors~\cite{Pour-Imani2016,Miller2019} and ages~\cite{Abdeen2022} and found a clear difference in those values that was consistent with predictions from density wave theory. However, they made conclusions based only on this finding, without~considering other arguments such as the profiles of angular shift and their connection to the rotation curve.
\par
In contrast to those mentioned papers, we not only found the existence of an age gradient in NGC~3686 predicted by the density wave theory, but~also demonstrated that its profile is described by the curve related to the velocity rotation curve (Fig.~\ref{fig:offset} at the top). Moreover, the~corotation radius, found as the location where the angular shift profile changes sign, is consistent with other measurements taken from the literature (Fig.~\ref{3686_consistency}, top panel). Furthermore, the~position of the corotation resonance was confirmed using another measurement obtained from kinematic data using the well-established method described in~\cite{Font2011}. Additionally, the~investigation of resonances (see Figure~\ref{3686_consistency}, bottom panel) related to the spiral pattern showed that the inner ultraharmonic resonance aligns with the ends of the bar (the beginning of the spirals). This also supports the density wave scenario related to the assumption that waves propagate within resonances. Despite some evidence appearing to be questionable, such as age gradient derived from modelled spirals instead of real ones, the~fact that all these corotation radius measurements and other arguments align well could provide a solid foundation for considering a density wave scenario as dominant in this galaxy.
\par
Based on our analysis, the~galaxy NGC~4321 is an interesting example of a grand-design spiral galaxy, whose spiral pattern may not be linked to a density wave that rotates at a constant speed. This object has been investigated in numerous studies in terms of corotation radius location. We have shown that CR measurements obtained using different methods are inconsistent (Fig.~\ref{fig:distr},middle panel), which can be explained by either a wide range taken by this resonance in the disk, or~by the absence of any corotation. As~for the existence of a stellar age gradient, the~results of different studies, including this paper, are quite ambiguous. Our analysis indicates a slight angular shift between the stellar patterns of different ages (see middle panel of Figures~\ref{fig:offset} and~\ref{NGC4321_age_gradient}), but~it does not change the direction. The~same picture was obtained by authors of~\cite{Sanchez-Gil2011}. This problem was also investigated in~\cite{Kendall2011} by analyzing the azimuthal offset between stellar and gas shocks. According to Figure~39 from that paper, for~most of the disk, the~angular shift is almost zero, which means that the stellar age gradient is negligible. In~addition, the~authors of~\cite{Ferreras2012} used the modelling of stellar populations to analyze the spatial distribution of stars having different ages, and~they also did not detect any significant offset expected in density wave scenario. However, there are some studies that have found an age gradient and identified a change in direction associated with the CR location. Those positions are marked by red cones in the middle panel of Figure~\ref{fig:distr}. The~location of 153~arcsec was taken from~\cite{Oey2003} and 87 and 114~arcsec from~\cite{Abdeen2020}. Note, these measurements are not consistent within their errors, which questions both the accuracy of the corotation position and the presence of an apparent age gradient.
\par
To summarize, NGC~4321 appears to have a slight stellar age gradient that is small enough to be detected in studies~\cite{Kendall2011,Ferreras2012}. Additionally, we did not observe a clear change in the direction of the age (or color) gradient. These findings suggest that the pattern speed likely changes in the same way as the angular velocity of the disk. This, in~turn, implies the presence of many corotation resonances, which can be seen in the distribution of CR measurements (Figure~\ref{fig:distr}).
\par
Finally, NGC~2403 has been found to be a good example of a galaxy that demonstrates what is known as ``mode-coupling'', which has been studied in detail in~\cite{Sygnet1988,Masset1997}. This means that a certain galaxy exhibits multiple patterns, each rotates with a different angular velocity, and~each occupies its own radial region. Unlike a single pattern, the~resulting of these waves can extend over a larger range of radii. Additionally, inner and outer patterns are expected to interlock through overlapping resonances, most commonly through the OLR and CR or ultraharmonic resonances~\cite{Rautiainen1999}. The~presence of these structures has been detected in real galaxies repeatedly. The~authors of~\cite{Meidt2008,Meidt2009} used the ``Radial Tremaine-Weinberg Method'', which involves changing the pattern speed with radius, and~demonstrated mode coupling for certain objects. Furthermore,~\cite{Buta&Zhang2009} found multiplicity in corotation radius positions for over 150 galaxies using the ``potential-density'' method. It is based on an analysis of the shift between the potential distribution and the density wave. The~authors of~\cite{Font2014a} also investigated interlocking resonances in real galaxies using the ``Phase Reversals Method'' (``F-B'' method) and found that 70\% of the galaxies studied had features of mode coupling. Unlike the mentioned studies, the~authors of~\cite{Beckman2018, Marchuk2024c, Kostiuk2024} used CR measurements obtained by at least two different methods to identify multiple patterns and their couplings. This approach allows not only verifying some measurements but also finding other CR positions that are impossible to obtain using other methods due to, for~example, limitations in observational data. In~our analysis, we applied three methods for corotation estimation and compared the obtained measurements with those from other studies. By~clarifying and verifying CR locations, we identified three structures rotating at different speeds (see Figure~\ref{2403_consistency}). In~addition, we found overlapping resonances as evidence of mode coupling. 
\par
Summarizing, we studied three spiral galaxies with regard to corotation radius locations. In~addition to the CR positions found in other studies, we applied three different methods to obtain independent measurements (see Table~\ref{tab2}). By~comparing gathered CR measurements which those obtained in this paper, we found that for NGC~3686, this resonance likely occurs at 43$_{-4}^{+7}$~arcsec, while for NGC~2403, in~line with previous findings~\cite{Marchuk2024c}, we found three possible corotation radii at 140$_{-12}^{+8}$, 298$_{-30}^{+13}$, and~385$_{-25}^{+30}$~arcsec. Further analysis of stellar age gradients and overlapping resonances confirmed the found CR locations and allowed us to conclude that the spiral structure of NGC~3686 likely follows a density wave scenario, and~NGC~2403 exhibits three patterns rotating at different angular velocities. Regarding NGC~4321, the~same analysis did not reveal any clear positions of corotation resonances. In~addition to the absence of a clear stellar age gradient, this implies that the nature of spiral structure of the galaxy may not be linked to density wave theory.
\par
This approach, based on examining the distribution of corotation radius measurements and applying additional observational tests, allows us to conclude that the spiral structures of all the galaxies we studied may have a distinct nature. The~origin of spirals in real galaxies is still a topic of debate due to discrepant results from different observational methods, which allow us to distinguish between dynamic and density wave scenarios. In~particular, in~our previous paper~\cite{Kostiuk2024}, we revealed inconsistencies in corotation radius measurements for a significant fraction of galaxies. A~number of objects with a reliably determined mechanism for spiral formation is very small. In~this paper, we demonstrate, for~the first time, a~surprising agreement between various pieces of evidence supporting one theory or another, making our results plausible. According to our analysis, the~galaxy NGC 3686, which does not have prominent and distinct spiral pattern, shows many arguments in favor of the density wave scenario. On~the other hand, NGC 4321, which demonstrate on the first sight grand-design spiral arms, is likely to have dynamic spirals. Therefore, our findings question the previous belief in a direct relationship between the origin of spiral structure and its appearance. What properties of galaxies lead to this difference in their features, demonstrating the approval of this or that scenario for the formation of spiral structures? Which mechanism is more prevalent among real galaxies? Further investigation of more objects using our approach may provide some clues to solving these~problems.

\vspace{6pt} 


\authorcontributions{Conceptualization, V.K. and A.M.; Methodology, V.K. and I.C.; Validation, A.M. and A.G.; Investigation, V.K.; Data Curation, A.M. and V.K.; Writing---Original Draft Preparation, V.K.; Writing---Review and Editing, A.M., A.G. and I.C.; Visualization, V.K. and I.C. All authors have read and agreed to the published version of the~manuscript.}

\funding{The study was conducted under the state assignment of Lomonosov State Moscow University. This research was funded by the Fund for the Development of Theoretical Physics and Mathematics ``BAZIS'' grant number~23-2-2-6-1.}

\dataavailability{The data underlying this article were accessed from publicly available cited references. The~dataset of corotation radius measurements is shared on \url{https://github.com/ValerieKostiuk/CRs\_dataset}~(accessed on 1 March 2024).} 


\conflictsofinterest{The authors declare no conflict of~interest.} 






\appendixtitles{yes} 
\appendixstart
\appendix
\section[\appendixname~\thesection]{Velocity Maps}
In this section, we present the velocity field maps (see the left column) obtained from H$\alpha$ observations, which were used in Section~\ref{sec:F-B}. By~applying the \textsc{BBarolo} tool with varying positional angle, we obtained the model of rotational velocities (in the central column). The~residual velocity maps, which were used for the Font-Beckman method, are shown in the right column.
\label{Ap:Vel_maps}
\begin{figure}[H]

\includegraphics[width=12.5cm]{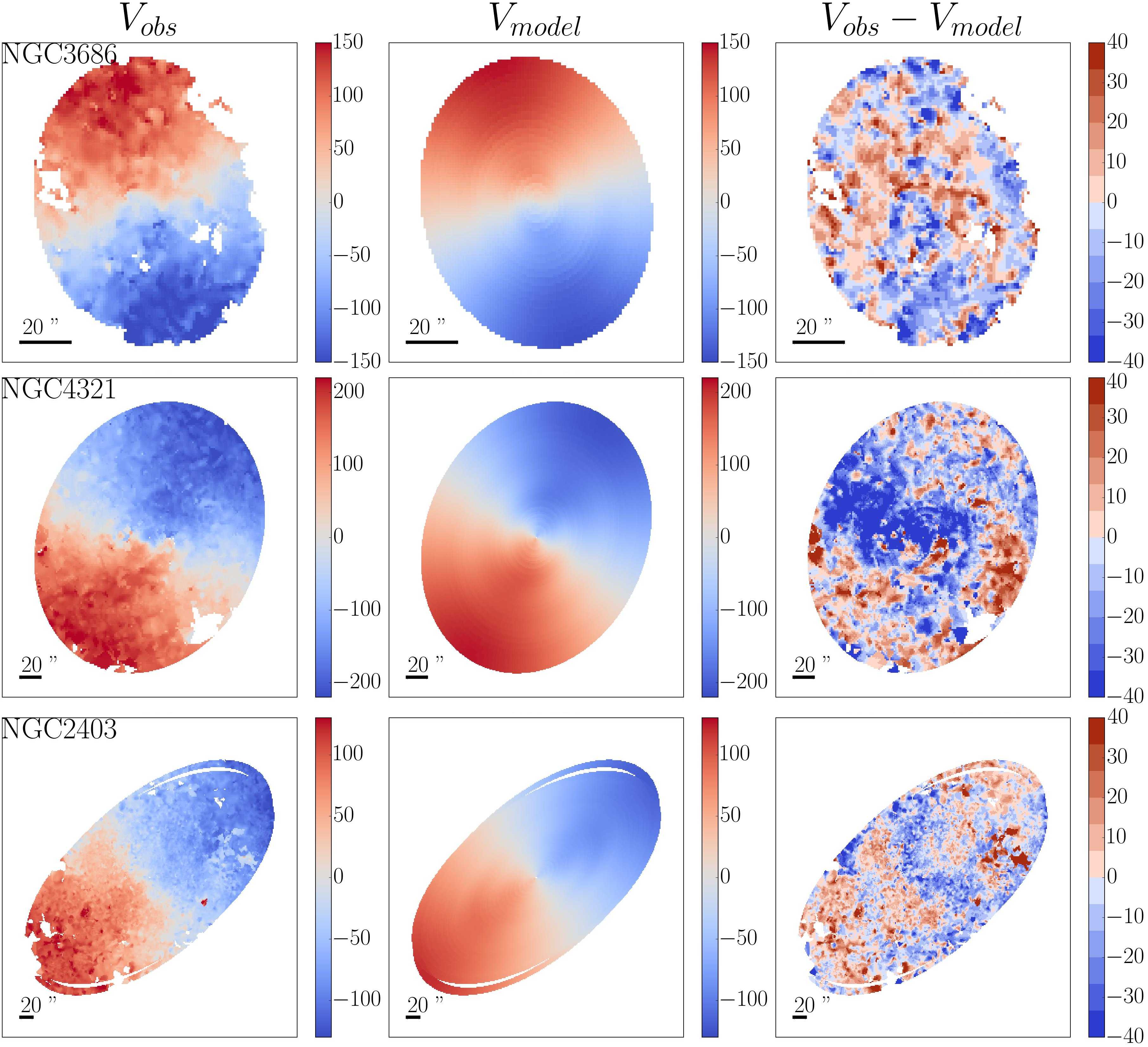}
\caption{\label{fig:vel_maps}Images of the observational velocity field (\textbf{left}), model of circular velocities obtained using \textsc{BBarolo} tool (\textbf{middle}) and the map of residual velocity (\textbf{right}). From~top to bottom: NGC~3686, NGC~4321, NGC~2403.}
\end{figure}
\unskip
\section[\appendixname~\thesection]{Decomposition with Spiral Arms}
\label{Ap:decomp}

We introduce a new method for measuring stellar age gradients (Section~\ref{sec:offset}). In~short, this method requires one to perform 2D photometric decomposition with spiral arms for IR and UV images, which gives an analytical approximation of the shape of spiral arms in both filters. Then, azimuthal offsets can be calculated directly from the parameters obtained which describe the shape of spiral arms. Our method for performing photometric decomposition with spiral arms is first presented in~\cite{Chugunov2024}. Here we will only provide an outline of it, and~we refer the reader to that paper for a detailed description and the reasoning of the entire process. The~method itself and the function describing surface brightness distribution in spiral arms were developed further in the following works:~\cite{Marchuk2024b} and in~\cite{2025arXiv250111670C}.

To analyze infrared spiral arms, we have utilized images from S$^4$G survey~\cite{Sheth2010} in 3.6 $\upmu$m band with a pixel size of $0.75''$. In~ultraviolet, we have used GALEX data from different surveys. For~NGC~2403 and NGC~4321 we utilized FUV band images, whereas for NGC~3686 we used NUV band data, because~FUV images for this galaxy all have too small exposition time. In~both bands, pixel size is $1.5''$.
Note that the difference in pixel size between the two images may affect the resulting angular offset shown in Figure~\ref{fig:offset}. However, this deviation does not exceed the error margins of the fitted profiles.

Photometric decomposition is the process of fitting some analytical model to the light distribution in a galaxy. The~model usually consists of a few components, such as bulge and disc, each modeled by some analytical function with a number of fitted parameters. To~perform decomposition, we use the IMFIT package~\cite{Erwin2015} which allows one to add custom functions; in our case this is a function that describes the 2D light distribution of an individual spiral arm. Our modified version is available online at~\url{https://github.com/IVChugunov/IMFIT_spirals} (accessed on 1 January 2024) 

We consider surface brightness to be a function of polar coordinates: galactocentric distance $r$ and azimuthal angle $\psi$ counted from the beginning of the spiral arm. These values are counted in the galactic plane. The~function can then be represented as a product of the following parts that are convenient to be considered separately:
\begin{equation}
    \label{eq:I_overall}
	I(r, \psi) = I_\parallel(r(\psi), \psi) \times I_\bot(r - r(\psi), \psi)\,,
\end{equation}
where $r(\psi)$ is a shape function that describes the overall shape of the spiral arm, $I_\parallel$ defines the light distribution along the arm and $I_\bot$ describes the surface brightness profile across the~arm. 

Our shape function allows one to reproduce sophisticated shapes of spiral arms. We call the curve $[r(\psi), \psi]$ a ridge-line of the spiral arm, as~for a given $\psi$, $r(\psi)$ is a radius where brightness is highest. We employ 3 varieties of our shape function; for each individual spiral arm, one is chosen based on its appearance, specifically the presence of visible sharp bends. These varieties are presented below:
\begin{equation}
	\label{eq:r_psi_simple}
	r(\psi) = r_0 \times \exp \left(\sum_{n=1}^4 k_n \psi^n\right)\,
\end{equation}
\begin{equation}
    \label{eq:r_psi_1b}
    \begin{cases}
    r(\psi) = r_0 \times \exp \left(\sum_{n=1}^3 k_n \psi^n\right),~~\psi < \psi_{b1}\\
    r(\psi) = r_{b1} \times \exp \left(\sum_{n=1}^3 l_n (\psi - \psi_{b1})^n\right),~~\psi \geq \psi_{b1}
    \end{cases}
\end{equation}
\begin{equation}
    \label{eq:r_psi_2b}
    \begin{cases}
    r(\psi) = r_0 \times \exp \left(\sum_{n=1}^2 k_n \psi^n\right),~~\psi < \psi_{b1}\\
    r(\psi) = r_{b1} \times \exp \left(\sum_{n=1}^2 l_n (\psi - \psi_{b1})^n\right),~~\psi_{b1} < \psi \leq \psi_{b2}\\
    r(\psi) = r_{b2} \times \exp \left(\sum_{n=1}^2 m_n (\psi - \psi_{b2})^n\right),~~\psi \geq \psi_{b2}
    \end{cases}
\end{equation}

Although they appear complex, these functions simply produce spiral arms with their pitch angles $\mu$ vary as polynomial functions of $\psi$. For~Function~\ref{eq:r_psi_simple} this is a simple polynomial of order 4, whereas Functions~\ref{eq:r_psi_1b} and~\ref{eq:r_psi_2b} are piecewise polynomial functions, allowing one to construct spiral arms with one or two sharp bends, if~needed. Therefore, $r_0$ describes the radius where the spiral arm begins, while $k_n$, $l_n$, $m_n$ are coefficients describing the shape of polynomial and $\psi_{b1}$ and $\psi_{b2}$ are azimuthal angles where bends are located. Finally, $r_{b1}$ and $r_{b2}$ are determined by the condition that the arm segments join each~other.

Then, the~function $I_\parallel$ basically defines the behavior of surface brightness along the $r(\psi)$ curve. $I_\parallel$ is the exponential function of the radius (with exponential scale $h_s$ and projected central brightness $I_\text{sp}$), modified by a truncation function of azimuthal angle $I_\text{tr}(\psi)$ which equals 1 in most part of the arm (between azimuthal angles $\psi_\text{gr}$ and $\psi_\text{cut}$), decreasing to zero near the beginning and end of the arm (the end is located at azimuthal angle $\psi_\text{end}$).
\begin{equation}
    I_\parallel(r, \psi) = I_\text{sp} \times e^{-r/h_s} \times I_\text{tr}(\psi)
\end{equation}
\begin{equation}
	I_\text{tr}(\psi) =
	\left\{\begin{array}{ll}
		3 \left(\frac{\psi}{\psi_\text{gr}}\right)^2 - 2 \left(\frac{\psi}{\psi_\text{gr}}\right)^3, & 0 \leq \psi < \psi_\text{gr}\\
		1, & \psi_\text{gr} \leq \psi < \psi_\text{cut}\\
		\frac{\psi_\text{end} - \psi}{\psi_\text{end} - \psi_\text{cut}}, & \psi_\text{cut} \leq \psi \leq \psi_\text{end}\,.
	\end{array}\right.\
\end{equation}

The function $I_\bot$ defines the surface brightness profile across the arm, more precisely along the line of constant $\psi$. Here we use $\rho = r - r(\psi)$, a~distance to the ridge-line, for~convenience, so at a given $\psi$ $I_\bot$ is an asymmetric Sersic function where $w_\text{loc}^\text{in/out}$ is the HWHM of the profile to the inner and outer direction, which is variable and depending on $\psi$. $n^\text{in/out}$ is a Sersic index, defining the shape of profile.
\begin{equation}
    \label{eq:I_bot}
	I_\bot^\text{in/out}(\rho, \psi) = \exp \left(-\ln(2) \times \left( \frac{|\rho|}{w_\text{loc}^\text{in/out}(\psi)} \right)^{\frac{1}{n^\text{in/out}}} \right)\,,
\end{equation}

$w_\text{loc}^\text{in/out}$ depends on $\psi$ in a following way. Here, $w_r$ is the total width of spiral arm at its middle in terms of $\psi$. $S$ is the skewness parameter that describes how wider or narrower the outer part is, compared to inner one. Finally, $\gamma_\text{in/out}$ define the rate of increase or decrease of the width along the spiral arm.
\begin{equation}
    \label{eq:w_loc}
	w_\text{loc}^\text{in/out}(\psi) = w_r \frac{1 \mp S}{2} \times \exp\left(\gamma_\text{in/out} \left(\frac{\psi}{\psi_\text{end}} - 0.5\right)\right)\,.
\end{equation}

In final form, the~function has 21, 24 or 25 parameters depending on which shape variety was used, for~a single spiral arm. Note that the number of free parameters was smaller, as~some parameters were fixed to some <<default>> value, especially in smaller arms. In~particular, $n^\text{in/out}$ in Equation~(\ref{eq:I_bot}) was fixed to 0.5, therefore perpendicular profile was Gaussian in all~cases.

For this work, our primary interest is the shape function of spiral arm. As~decomposition yields the full set of best-fitting model parameters, the~exact $r(\psi)$ in UV and in IR can be calculated directly, and~then azimuthal offsets as a function of radius~obtained.

In Figure~\ref{fig:decomp_models} we show a decomposition models of all three galaxies in~infrared.

\begin{figure}[H]

\centering
\includegraphics[width=13.5cm]{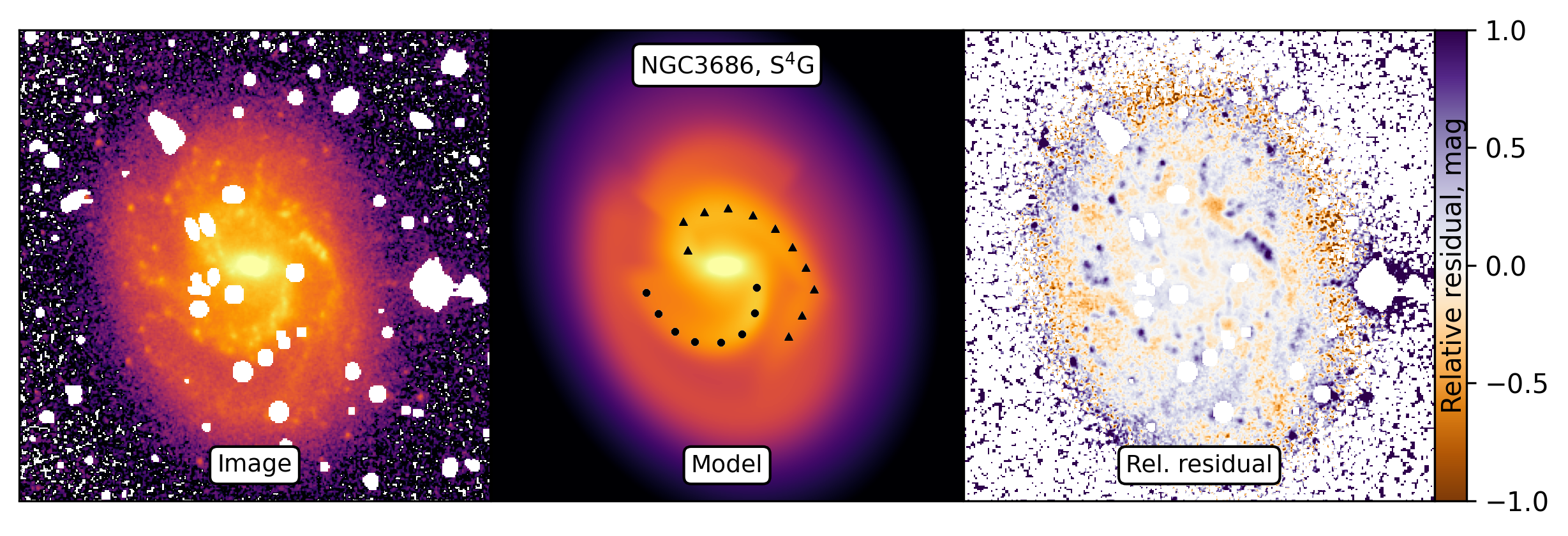}
\includegraphics[width=13.5cm]{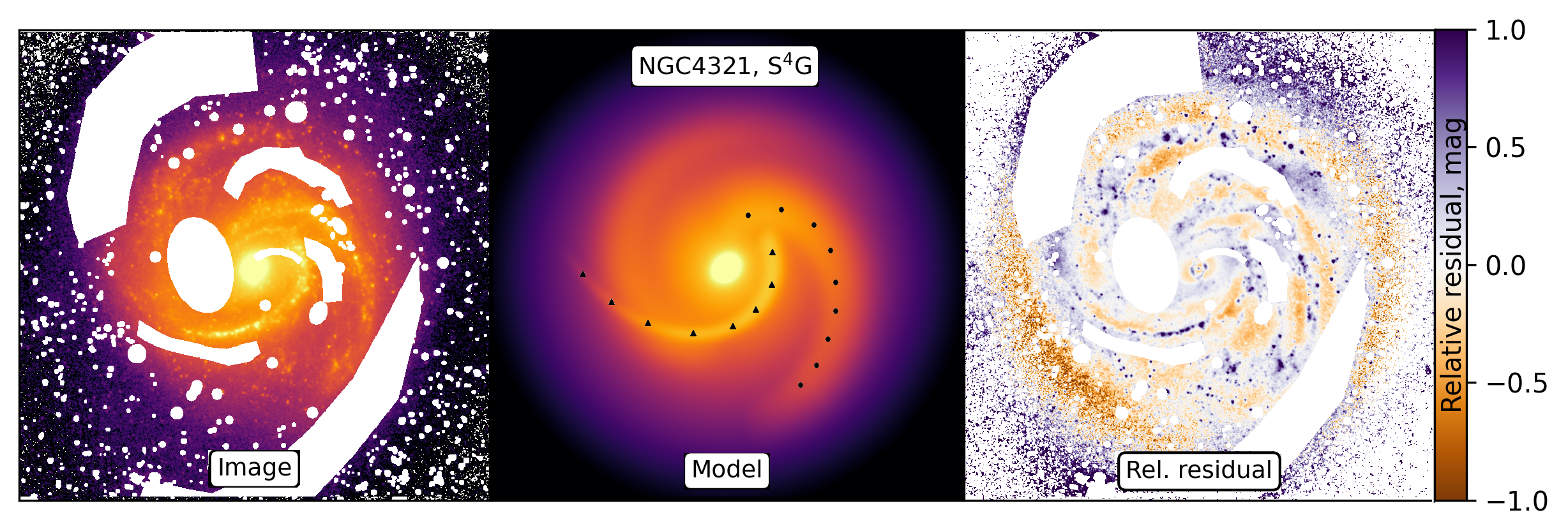}
\caption{\textit{Cont.}}
\end{figure}
\begin{figure}[H]\ContinuedFloat
\includegraphics[width=13.5cm]{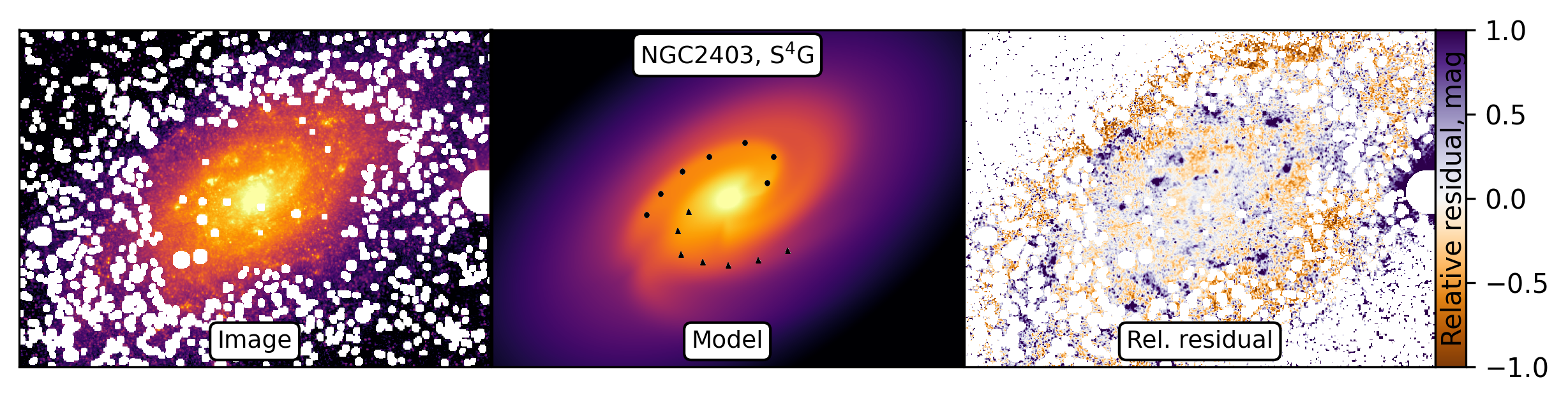}
\caption{\label{fig:decomp_models} Images of modelled galaxies in 3.6 $\upmu$m (\textbf{left}), corresponding decomposition model (\textbf{middle}) and relative residuals (\textbf{right}). From~top to bottom: NGC~3686, NGC~4321, NGC~2403. In~the middle panel, spirals are marked with circles and triangles, demonstrating which arm was used in the analysis shown on Figure~\ref{fig:offset}. }
\end{figure}

\begin{adjustwidth}{-\extralength}{0cm}
\printendnotes[custom] 

\reftitle{References}




\PublishersNote{}
\end{adjustwidth}
\end{document}